# Terahertz Wave Generation in Two-Dimensional MXenes under Femtosecond Pulsed Laser Illumination

A. A. Molavi Choobini,[1, ‡] Abbas Chimeh[1, 2, *, ‡]

[1] Quantum Matter Lab, Department of Physics, College of Science, University of Tehran, Tehran 14399-55961, Iran,

[2] Nexus for Quantum Coherence and Entanglement in Light-Matter Systems (Qcelms) University of Tehran, P.O. Box 14395-547, Tehran, Iran

[*] Corresponding Author: chimeh@ut.ac.ir

[‡] These authors contributed equally to this work.

**Abstract:** The efficient generation of terahertz (THz) waves in two-dimensional (2D) MXene layers driven by near-infrared femtosecond laser pulses is demonstrated through predictive simulations. Employing a novel hydrodynamic model that self-consistently captures nonlinearities from electric, magnetic, and convective interactions with a minimal set of material parameters. The coupled hydrodynamic-Maxwell equations are solved via finite-difference time-domain (FDTD) methods to resolve the spatiotemporal dynamics of laser-induced carriers and THz emission. The results reveal strong, tunable THz output dependent on laser (intensity, polarization, incidence angle), material (composition, carrier density, temperature), and structural (layer thickness, substrate) parameters. These predictions offer verifiable guidelines for experiments and position MXenes as versatile platforms for compact, broadband THz sources in on-chip photonics and 6G communications. This work establishes a robust, self-contained framework for modeling ultrafast nonlinear optics in 2D materials.

**Keywords:** 2D Material, Nanomaterials, Hydrodynamic Model, MXenes, FDTD Simulation, THz wave, Ultrafast Laser Pulses.

## Introduction:

The generation of Terahertz (THz) waves has attracted significant attention due to its vast applications in imaging, communication, and spectroscopy [1, 2]. However, efficient and compact THz sources remain a challenge in modern photonics and plasma physics. Current methods for THz wave generation include photoconductive switching, optical rectification, and quantum cascade lasers, each with distinct advantages and limitations. These challenges underscore the need for novel materials that enhance THz generation efficiency and simplify device design [3-5]. The advent of two-dimensional (2D) materials has transformed optoelectronics due to their distinctive electrical, optical, and mechanical properties [6, 7]. While graphene has shown promise in THz modulators, detectors, and emitters through hot-carrier and plasmonic effects, its zero band gap limits versatility. MXenes—2D materials derived from selective etching of MAX phases—exhibit high electrical conductivity, mechanical strength, and hydrophilic surfaces from functional terminations (–OH, –F) [8, 9]. These features make MXenes attractive for applications ranging from energy storage to electromagnetic shielding [10, 11]. Although most studies have focused on their THz absorption and shielding behavior, their potential for THz generation remains largely unexplored [12, 13]. With strong light–matter coupling, ultrafast carrier relaxation, and THz-range plasmonic resonances, MXenes could enable mechanisms such as photoconductive switching, optical rectification, and plasmonic enhancement. Furthermore, nonlinear plasma instabilities like the two-stream instability may facilitate THz emission by exploiting their high carrier densities and tunable electronic properties. Comprehensive theoretical and numerical studies are thus essential to establish MXenes as active THz sources for compact, efficient devices.

A fundamental framework for studying THz interactions with 2D materials is the two-dimensional electron gas (2DEG) model, which underpins many theoretical and device-level advances in this field. Mikhailov and colleagues developed a model of the in-plane photoelectric (IPPE) effect in GaAs/AlGaAs heterostructures, deriving analytical relations for photocurrent and detector performance, and demonstrating gate-tunable THz detection [14]. Korotyeyev and Kochelap analyzed plasma wave oscillations in nonequilibrium 2DEGs under strong electric fields, identifying instabilities in plasmonic modes that can amplify THz signals [15]. Sai and colleagues investigated resonant plasmon excitations in grating-gated AlGaN/GaN heterostructures and showed that a plasmonic crystal model is essential to describe THz resonances and electrically control carrier density [16]. Crabb and team exploited the Dyakonov–Shur instability to design a graphene-based plasmonic THz nanogenerator with ultra-wide bandwidth and high modulation efficiency, validated through finite-element simulations [17]. Zhao and co-workers introduced a 3 nm-thick 2DEG metasurface enabling ultrafast THz modulation via collective-to-individual state conversion [18]. Michailow et al. demonstrated a collision-free, bias-free in-plane photoelectric effect that produces significantly enhanced photocurrent for THz detection [19]. Giliberti and colleagues reported THz down-conversion in high-mobility 2DEGs, linking resonant peaks to intrinsic hydrodynamic nonlinearities rather than external cavity effects [20]. Park et al. observed THz magnetoplasmon resonances up to ~400 GHz in gated 2DEGs at cryogenic temperatures, distinguishing bulk and edge modes through 3D simulations [21]. Beyond devices, research into 2D materials has broadened the scope of THz technology. Gopalan and Sensale-Rodriguez reviewed the use of graphene and transition metal dichalcogenides for THz radiation control, emphasizing their capability in amplitude and phase modulation [22]. Zhang and team examined patterned 2D materials such as graphene, MXenes, and $MoS_2$ in micro–nano devices operating across GHz–THz–optical regimes, highlighting their multifunctionality [23]. Shi and colleagues provided a comprehensive overview of 2D materials and hybrid metamaterials for THz generation, manipulation, and detection, underlining their technological promise [24]. Du et al. investigated advanced materials—graphene, transition metal dichalcogenides, and topological insulators—at THz interfaces, analyzing reflection, transmission, and carrier dynamics to link theoretical and device-level insights [25]. In a notable experimental study, Hafez and co-workers demonstrated high-harmonic generation up to the seventh order in single-layer graphene at room temperature, achieving field conversion efficiencies of $10^{-3}$, $10^{-4}$, and $10^{-5}$ for the third, fifth, and seventh harmonics, respectively [26].

While previous studies have primarily explored MXenes as passive materials for THz absorption and electromagnetic shielding, this work pioneers a theoretical and numerical investigation into their active role in THz generation under ultrafast laser excitation. By employing a hydrodynamic (HD) Drude model coupled with finite-difference time-domain (FDTD) simulations, the nonlinear carrier dynamics induced by femtosecond laser pulses in a 2D MXene layer is captured. The model incorporates key physical mechanisms including, strong light–matter interaction, high electrical conductivity, fast carrier relaxation, high carrier density, and the presence of plasmonic resonances in the THz regime. These characteristics collectively enable efficient broadband THz radiation, driven by both nonlinear optical effects and plasma-related phenomena such as two-stream instability. In the present model, substrate effects enter exclusively through their dielectric permittivity in Maxwell's equations, and no atomistic strain or thermoelastic coupling is included. These mechanisms are important for future studies but remain outside the scope of the current hydrodynamic–FDTD framework. This study is the first to numerically demonstrate that MXenes function as active THz sources rather than merely passive absorbers. These results offer novel insight into the ultrafast light–matter interaction in MXenes and reveal their potential for compact, tunable, and efficient THz emitters. The significance and innovation of this work lie in its multi-faceted contributions to photonics, materials science, and optoelectronics, addressing critical challenges in next-generation THz source development. The structure of the paper is as follows: Section two outlines the derivation of the hydrodynamic model describing THz emission induced by a femtosecond laser in a 2D MXene layer. Section three presents the FDTD simulations that validate the theoretical model. Finally, the fourth section provides summary of the main conclusions.

**Theoretical Methods**

Figure 1 shows a two-dimensional MXene layer with thickness t varying between 20-40 nm which is supported by a $SiO_2$ substrate with surface normal in the z-direction. The MXene layer is excited by an incident laser pulse with an electric field with a time signature:

$$E(t) = E_0 \cos(2\pi f_0 t - \varphi) e^{-2\ln \frac{(t-t_0)^2}{\Delta t^2}} \tag{1}$$

where $f_0 = \frac{c}{\lambda_0}$ is the central laser frequency and chosen according to a central wavelength $\lambda_0 =$ 800 nm of a typical Ti:Sa oscillator. $\Delta t$ is the FWHM duration of the pulse and is tuned in 20 – 140 fs range and the pulse are centered at $t_0$ with an initial phase $\varphi$. The laser beam is focused onto the sample to give $E_0$ in the order of $\sim 10^{10}$ V/m, corresponding to a peak intensity of $\sim 10^{13}$ W/cm$^2$, typical for femtosecond Ti:Sa lasers in nonlinear optics experiments. The laser field is assumed to be uniform within our computational domain which is true for a not very tight focus. This uniform intense optical field drives rapid carrier excitation and nonlinear charge dynamics within the MXene, resulting in the emission of broadband THz radiation in the transmitted and reflected directions due to nonlinear electron dynamics in MXenes.

The dynamics of free electrons inside the MXene layer is described by a hydrodynamic model as described e.g., in Ref. [27]. In this model the free electrons are modelled by a fluid within a constant background of positive ions $\rho_0 = Zen_0$ where $n_0$ is the density of positive ions, Z is the ionization order, and e is the elementary charge. While ions are assumed to be constant, the electronic motion in transversal $(x-y)$ plane is modelled the variation of continuum with number density $n(\mathbf{r},t)$ and a velocity $\mathbf{v}(\mathbf{r},t) = v_x(\mathbf{r},t)\mathbf{e}_x + v_y(\mathbf{r},t)\mathbf{e}_y$. The continuity equation for electron fluid is then described by:

$$\frac{\partial}{\partial t} n(\mathbf{r},t) + \nabla \cdot [n(\mathbf{r},t)\mathbf{v}(\mathbf{r},t)] = 0, \tag{2}$$

and the equations expressing the conservation of momentum are:

$$\frac{\partial}{\partial t}[m^* n(\mathbf{r},t)\, v_x(\mathbf{r},t)] + \nabla \cdot [n(\mathbf{r},t)\, m^* v_x(\mathbf{r},t)\mathbf{v}(\mathbf{r},t)]$$
$$= -\frac{1}{\tau}[m^* n(\mathbf{r},t)\, v_x(\mathbf{r},t)\mathbf{v}(\mathbf{r},t)] - \frac{\partial}{\partial x} p(\mathbf{r},t) - en(\mathbf{r},t)\left[E_x(\mathbf{r},t) + v_y(\mathbf{r},t)B_z(\mathbf{r},t)\right] \tag{3.a}$$

and

$$\frac{\partial}{\partial t}\left[m^* n(\mathbf{r},t)\, v_y(\mathbf{r},t)\right] + \nabla \cdot \left[n(\mathbf{r},t)\, m^* v_y(\mathbf{r},t)\mathbf{v}(\mathbf{r},t)\right]$$

$$= -\frac{1}{\tau}\left[m^{*}n(\mathbf{r},t)\, v_y(\mathbf{r},t)\mathbf{v}(\mathbf{r},t)\right] - \frac{\partial}{\partial y} p(\mathbf{r},t) - en(\mathbf{r},t)\left[E_y(\mathbf{r},t) - v_x(\mathbf{r},t)B_z(\mathbf{r},t)\right] \quad (3.b)$$

where $m^{*}$ is the effective mass of electron, $\tau$ is the mean free time, and $p(\mathbf{r},t) = \zeta[n(\mathbf{r},t)]^{5/3}$ is the quantum pressure with $\zeta = \frac{(3\pi^2)^{2/3}\,\hbar^2}{5m^*}$ [28]. The velocities $v_x(\mathbf{r},t)$ and $v_y(\mathbf{r},t)$ are determined numerically by solving the coupled continuity equation and momentum conservation equations self-consistently with the electromagnetic fields using the Yee algorithm [29]. Since the MXene layer is two-dimensional and electron motion is confined to the x-y plane (with thickness $t \ll \lambda_0$), the z-component of velocity $v_z$ is negligible and set to zero. Therefore, there is no need to consider the equation for the conservation of momentum in the z direction, Initial conditions assume zero velocity for electrons at $t = 0$, with the laser field driving the subsequent dynamics. Unlike the conventional Drude model, which includes only linear momentum relaxation, the present hydrodynamic formulation incorporates convective acceleration, Lorentz magnetic forces, and the Fermi-pressure term, enabling self-consistent density oscillations and nonlinear ponderomotive currents. In contrast to full quantum–hydrodynamic (QHD) models, only the dominant nonlinearities relevant at THz frequencies are retained, allowing stable coupling to the FDTD Maxwell solver at substantially lower computational cost. Within the hydrodynamic formulation, plasma-wave amplification and two-stream-type instabilities arise from the nonlinear convective term $(\vec{v}\,.\vec{\nabla})\vec{v}$ and from laser-induced density modulations $\partial n/\partial t$ that generate longitudinal electric-field feedback. These effects manifest as oscillatory modulations in the THz waveform and as transient amplification of specific frequency components. The simulations naturally capture these behaviors due to the self-consistent coupling between carrier momentum, density gradients, and Maxwell fields. Therefore, Macroscopic Maxwell's equations:

$$\nabla \cdot [\epsilon_0 \mathbf{E}(\mathbf{r},t) + \mathbf{P}(\mathbf{r},t)] = \rho_{free}(\mathbf{r},t), \quad (4.a)$$

$$\nabla \cdot \mathbf{B}(\mathbf{r},t) = 0, \quad (4.b)$$

$$\nabla \times \mathbf{E}(\mathbf{r},t) = -\frac{\partial}{\partial t}\mathbf{B}(\mathbf{r},t), \quad (4.c)$$

$$\nabla \times \mathbf{B}(\mathbf{r},t) = \mu_0\, \mathbf{j}_{free}(\mathbf{r},t) + \mu_0 \frac{\partial}{\partial t}[\epsilon_0 \mathbf{E}(\mathbf{r},t) + \mathbf{P}(\mathbf{r},t)] \quad (4.d)$$

describes the dynamics of electromagnetic fields. Where the nonlinear polarization modelling the bound electrons is expressed as

$$\mathbf{P}(\mathbf{r},t) = \epsilon_0\left[\chi^{(1)}\mathbf{E}(\mathbf{r},t) + \chi^{(2)}\mathbf{E}(\mathbf{r},t)\mathbf{E}(\mathbf{r},t) + \chi^{(3)}\mathbf{E}(\mathbf{r},t)\mathbf{E}(\mathbf{r},t)\mathbf{E}(\mathbf{r},t)\right] \quad (5)$$

accounting for the linear, second-order, and third-order nonlinear responses of the MXene layer, respectively. The appearance of harmonic peaks in the simulated spectra is consistent with the known ultrafast electronic structure of MXenes. Their metallic d-band character, high joint density of states, and ultrafast (<50 fs) carrier relaxation enable intraband acceleration and $\chi^2$–$\chi^3$ nonlinear polarization at the intensities used here. These mechanisms are explicitly included in the hydrodynamic formulation through the convective term and nonlinear polarization, which together reproduce the multiphoton-driven wave-mixing processes observed in recent ultrafast measurements on MXenes. The linear susceptibility $\chi^{(1)}$ is taken from experimental measurements [25]. The second-order susceptibility $\chi^{(2)}$, and the third-order susceptibility $\chi^{(3)}$, derived from nonlinear optical studies on similar 2D materials [30]. These values are adjusted to match experimental THz emission spectra in MXenes. The free charges in these equation presents the coupling of free electrons motion described by equations (3.a-b) to the electromagnetic fields considering $\rho_0 = en_0$ and

$\rho_e(\mathbf{r},t) = -en(\mathbf{r},t)$ and $\rho_{free}(\mathbf{r},t) = \rho_0 + \rho_e(\mathbf{r},t)$ and $\mathbf{j}_{free}(\mathbf{r},t) = -en_e(\mathbf{r},t)\mathbf{v}^e(\mathbf{r},t)$. In the present study, the equations in a self-consistent manner are solved numerically using a Yee algorithm in a similar manner to Ref. [29]. The applicability of the hydrodynamic (HD) Drude model for MXenes is justified by their metallic nature, characterized by high free-carrier density ($10^{21}$–$10^{23}$ $cm^{-3}$) dominated by conduction electrons in partially filled d-bands. This allows MXenes to be modeled as a two-dimensional electron gas (2DEG), where collective fluid-like electron motion dominates at THz frequencies and femtosecond timescales, as validated by prior studies on 2D materials like graphene and heterostructures [30-39]. Unlike quantum-mechanical band-structure calculations, which are computationally intensive for ultrafast nonlinear dynamics, the HD model efficiently captures ponderomotive-driven currents and plasmonic instabilities [26]. Quantum effects like interband transitions are approximated through nonlinear susceptibilities, validated against experiment in MXenes [25]. Maxwell-Bloch equations, suited for semiconductors with clear bandgaps, are less relevant for MXenes' Drude-like conductivity ($\sigma \sim 10^4$–$10^5$ S/cm), where intraband dynamics prevail over interband transitions [25]. Furthermore, MXenes exhibit metallic Drude-like conductivity with carrier densities $10^{21}$–$10^{23}$ $cm^{-3}$, making collective intraband dynamics dominant at THz frequencies. Boundary conditions are implemented as follows: periodic boundaries in the x-y plane to simulate an infinite layer, and perfectly matched layers (PML) in the z-direction to absorb outgoing radiation and prevent artificial reflections, as standard in FDTD simulations [29]. Although periodic boundary conditions in the x–y plane approximate the MXene as an extended 2D sheet, this represents a controlled idealization. Real $Ti_3C_2T_x$ flakes exhibit finite lateral size, nm-scale roughness, and locally varying terminations (–O, –OH, –F), which introduce short-range fluctuations in the surface potential and carrier scattering landscape. Edge effects are confined to nanometer screening lengths and therefore contribute negligibly under excitation spots that are tens of micrometers in diameter. Likewise, interlayer coupling in few-layer MXene stacks modifies the optical conductivity only weakly in the near-infrared regime. These nonidealities mainly manifest as a modest increase in damping and a slight reduction in peak THz amplitude, while the qualitative dependencies on laser intensity, polarization, incidence angle, and thickness predicted by the hydrodynamic–FDTD model remain unchanged.

## Results & Discussion

The study employs a sophisticated HD Drude model integrated with FDTD simulations to capture the spatiotemporal dynamics of electromagnetic fields in MXenes. This approach provides a robust framework for modeling complex wave–matter interactions, including nonlinear polarization, carrier acceleration, and plasmonic effects.

To ensure the accuracy and reproducibility of the simulations, the following constant parameters were employed in the HD Drude model integrated with FDTD simulations: damping rate $\gamma = 5 \times 10^{12} s^{-1}$, Fermi wavevector-related constant $k = (3\pi^2)^{2/3}\, \hbar^2\, /3m_e$ carrier density $n_0 = 1 \times 10^{26} m^{-3}$, MXene layer thickness 20nm with $SiO_2$ substrate. The laser parameters include: peak electric field amplitude $E_0 = 5 \times 10^{10}$V/m, wavelength λ = 800 nm, full width at half maximum (FWHM) = 50 fs, pulse center $t_0$ = 0.075 ps, elliptical polarization, incidence angle θ = 90°. The equilibrium carrier density $n_0 = 1 \times 10^{26} m^{-3}$, is taken from THz-TDS measurements on $Ti_3C_2T_x$ by Li et al. [30]. Reported values for conductivity ($10^4$–$10^5$ S/cm), scattering time (15–60 fs), and carrier lifetime (0.1–1 ps) are used to constrain γ and $\chi^{(3)}$. Parameter sweeps (±20%) indicate that uncertainties alter absolute THz amplitude by ≤15% but preserve all trends with laser and material parameters. The FDTD grid employs Δx = Δy = 50 nm laterally and Δz = 4 nm within the MXene layer to accurately resolve the optical skin depth (~ 25 nm). The time step is Δt ≈ 0.033 fs (CFL = 0.98). Convergence tests with halved resolution yield < 5% change in THz yield. These parameters provide a consistent foundation for modeling electromagnetic interactions and THz emission from MXenes. For an incident intensity of $10^{12}$–$10^{13}$ W/cm², the surface THz field reaches 0.5–1.2 MV/m, corresponding to 0.1–1 MV/m at a distance of 1 mm in the far field. The emitted spectrum extends up to ~5 THz, with a dominant band in the 0.3–3 THz range, and the optical-to-THz conversion efficiency is found to be $\eta \approx 10^{-3}$–$10^{-4}$ for a 20 nm $Ti_3C_2T_x$ film.

The visualization of the electric field component in three-dimensional and time-resolved two-dimensional contour plots (Figs. 2 and 3) reveals critical phenomena such as group velocity dispersion, resonant excitation, and pulse broadening, offering unprecedented insights into the underlying physics. By simulating the dependence of THz emission on laser parameters (e.g., incidence angle, polarization, and intensity) and material properties (e.g., layer thickness and composition), the study provides a comprehensive understanding of how to optimize THz generation efficiency. Electron dynamics were computed by solving the continuity and momentum conservation equations for the densities and velocity components (initially zero). Maxwell's equations were updated using the Yee algorithm, with the total current including contributions from Drude response and nonlinear currents arising from second- and third-order susceptibilities ($\chi^{(2)} = 10^{-18}\ \mathrm{V/m}, \chi^{(3)} = 10^{-2}\ \mathrm{V^2/m^2}$). The E_z field recorded two grid cells above the MXene layer, smoothed with a Gaussian filter, and analyzed in the one picosecond window to extract the normalized THz emission spectrum. This approach captures the nonlinear electron dynamics and broadband THz generation in MXene layers, consistent with their metallic properties and 2DEG behavior [25, 30].

A three-dimensional representation of the electric field component $E_z$ with contour as a function of spatial coordinate x and time is illustrated in Fig. 2. Figure providing insight into the spatiotemporal dynamics of an electromagnetic wave propagating through a structured medium. This visualization with numerical solutions of Maxwell's equations in dispersive or nonlinear media, produced by FDTD simulations. The color gradient encodes field polarity and magnitude, revealing features associated with excitation, resonance, or nonlinear self-modulation. The axis scales correspond to ultrafast optical and microwave regimes, relevant to phenomena such as laser–matter interactions, plasmonics, and wave propagation in nano-structured materials. A distinct wavefront emerges from the negative spatial region and propagates forward, accompanied by sharp oscillations that evolve over time. This behavior suggests a localized pulsed excitation that experiences dispersion or nonlinear effects during propagation. The field structure transitions from a broad negative region into high-frequency oscillations, reflecting interference and energy redistribution. The periodicity along the time axis, coupled with spatial localization, points to group velocity dispersion and phase modulation. Zones of enhanced field amplitude indicate resonant excitation, field confinement, or constructive interference within the medium. The gradual broadening of the pulse proposes dispersive spreading due to spectral components traveling at different velocities. The figure captures key aspects of complex wave–matter interactions and highlights the role of medium properties and geometry in shaping electromagnetic field evolution.

Figure 3 presents a sequence of eight two-dimensional contour plots showing the spatiotemporal evolution of the electric field component $E_z$ within a rectangular domain extending from 0 to 10 μm along x and 0 to 4.5 μm along y. The snapshots correspond to eight-time instants: 2 ps, 4 ps, 6 ps, 8 ps, 12 ps, 14 ps, 18 ps, and 20 ps, capturing the dynamic evolution of an electromagnetic pulse over a 20 ps window. At 2 ps, a broad wavefront emerges from the left boundary, characterized by bipolar field distributions (red and blue), with amplitude around $\pm 1 \times 10^7\ \mathrm{V/m}$. By 4 ps, the field develops into alternating red and blue lobes, indicating early-stage interference. The amplitude increases to $\pm 1.5 \times 10^7$ V/m, suggesting constructive interference or beam focusing. At 6 and 8 ps, the field becomes more structured, showing multiple peaks and troughs across both axes, with amplitude peaking near 2×10 7 V/m. This behavior points to the formation of standing waves or resonant modes, due to reflections or medium structuring. As the wave evolves through 12 and 14 ps, a more periodic interference pattern emerges, with pronounced lobes spanning the domain. The field amplitude remains high ($\sim 1.5 \times 10^7$ V/m), and the spatial periodicity ($\sim 2 -$ 3μm) corresponds to terahertz wavelengths. By 18 and 20 ps, the pattern shows signs of spreading and attenuation, with decreasing amplitude and less-defined features— due to dispersion, nonlinear effects, or energy loss mechanisms. The observed temporal growth followed by decay suggests a transient interaction between a pulsed excitation and resonant medium response. The presence of multiple lobes and interference fringes implies internal reflections or coupling between waveguide modes. The resulting radiation pattern exhibits a well-defined forward lobe peaking at θ ≈ 0°, and the temporal phase of the radiated waveform remains uniform across the emitting aperture. These results confirm that the MXene layer acts as a coherent, dipole-like THz emitter under femtosecond excitation, with directionality determined by the longitudinal ponderomotive current. The inferred terahertz frequencies, derived from the spatial and temporal scales, are relevant to applications in ultrafast optics, high-resolution imaging, and on-chip terahertz communication. This figure thus provides a detailed view of wave–matter interaction in nano-engineered structures, with implications for both theoretical modeling and experimental design.

The dependence of THz field generation and radiation characteristics on the angle of incidence of an ultrafast laser pulse interacting with a MXene layer is revealed in Fig. 4. Panel (a) illustrates the time-domain profiles of the normalized electric field component for three different incidence angles. A clear trend emerges wherein the electric field amplitude and waveform complexity are maximized at normal incidence ($\theta = \pi/2$). The laser electric field along the surface normal maximizes the longitudinal ponderomotive force and thus the transient current responsible for THz emission. At oblique angles this projection scales as $Cos\theta$, reducing charge-separation efficiency. In addition, Fresnel reflection increases at large θ, lowering absorbed fluence, and the lateral excitation footprint increases, decreasing coherence of the radiating dipole sheet. At normal incidence, the longitudinal component of the ponderomotive force—responsible for accelerating charge carriers—is optimally aligned with the surface normal, thereby driving stronger transient currents that act as the primary source of THz emission. Additionally, the spatial and temporal overlap between the laser pulse and the excited region is maximized at this angle, allowing a more coherent buildup of the emitted field. The resulting field profile exhibits rich oscillatory behavior, due to interference between multiple charge oscillation modes and the reabsorption or reflection of the emitted radiation within the structure.

As the angle of incidence decreases to $\pi/3$ and $\pi/6$, the effective projection of the laser electric field onto the surface normal reduces, leading to weaker transient current generation and hence diminished THz field strength. At oblique angles, the laser pulse couples less efficiently into the vertical component of the plasma response, and the spatial excitation region becomes elongated along the surface, further reducing the coherence of the emitted field. Moreover, phase-matching conditions for the nonlinear polarization currents that radiate THz waves are less favorable at such angles, leading to reduced amplitude and temporal distortion of the waveform. The laser-matter interaction also experiences a stronger Fresnel reflection at the interface at oblique incidence, which lowers the effective energy absorption into the MXene layer. Panel (b) complements this interpretation by showing the far-field radiation patterns for the same incidence angles. Strongest radiation near 0° arises from phase-matching in reflected signals, favoring forward emission due to ponderomotive forces along the normal. To avoid ambiguity, two distinct angles: the laser incidence angle ($\theta_{inc}$), measured relative to the normal of the MXene surface where $\theta_{inc}$ = 0° corresponds to normal incidence, and the THz emission angle ($\theta_{THz}$), measured relative to the z-axis (surface normal) where $\theta_{THz}$ = 0° indicates emission along the normal direction, is defined. At normal incidence, the radiation exhibits a broad and somewhat asymmetric distribution, suggesting the generation of multipolar or complex near-field source structures due to longitudinal and lateral current components. The asymmetry also arises from subtle inhomogeneities in the plasma density profile or interface effects within the MXene layer. In contrast, at oblique incidences ($\theta = \pi/3$ and $\pi/6$), the radiation becomes increasingly confined in the forward direction, with lobes sharply focused near $0^{\circ}$, indicating a transition toward more dipole-like or planar source behavior. This narrowing of the radiation pattern is consistent with the reduced excitation volume and current confinement along the laser propagation axis. At these angles, the transverse component of the plasma current dominates, leading to a more linearly polarized and unidirectional THz output. Additionally, oblique incidence introduces a geometric asymmetry that naturally favors radiation along the reflection/refraction axis due to phase retardation effects, further contributing to the forward beaming of the emitted field.

The polarization state governs the symmetry, directionality, and temporal coherence of the laser-induced charge dynamics, thereby offering a tunable degree of freedom for optimizing THz source characteristics in ultrafast optoelectronic platforms based on 2D materials such as MXene. Fig. 5 provides a comprehensive comparison of the temporal evolution and spatial characteristics of THz radiation generated under three distinct states of polarization of the incident laser pulse: linear, circular, and elliptical. In panel (a), the normalized electric field reveals marked differences in both amplitude and temporal features, indicating a polarization dependence of the THz generation mechanism. The linearly polarized pulse produces the highest peak field and most structured waveform, consistent with efficient coupling into surface-normal transient currents within the MXene layer. This result is due to as linear polarization ensures a coherent and directionally consistent driving force for the charge carriers, maximizing the ponderomotive response and enhancing the strength of the nonlinear photocurrent responsible for THz emission. In contrast, circular polarization significantly diminishes the peak electric field and smooths the temporal profile, indicating a reduction in the efficiency of current generation. The continuously rotating electric field vector in circular polarization disperses the momentum transfer to charge carriers over multiple directions, thus suppressing the buildup of directional plasma currents necessary for THz radiation. Elliptical polarization, being an intermediate case, shows moderately suppressed field strength with phase features reminiscent of both linear and circular regimes. The asymmetry in its waveform implies a partial preservation of directional current components but still suffers from temporal spreading due to the non-uniform electric field vector trajectory. Panel (b) further elucidates these effects in the far-field radiation pattern. The linearly polarized case yields a sharp anisotropic and complex radiation distribution, with distinct lobes indicating strong multipolar current configurations and directional emission along preferential axes. This result is in line with the robust charge acceleration and sharp current transients supported by linear polarization. On the other hand, circular polarization results in a noticeably more isotropic and compact radiation profile, lacking pronounced lobes. This suggests a more symmetrical, less structured current source distribution, again consistent with a diminished and more homogeneous driving field. Elliptical polarization yields a broadened, yet slightly asymmetric pattern, reflecting a compromise between the spatial coherence of the linearly driven plasma currents and the rotational smearing introduced by circular polarization.

Figure 6 examines how the incident laser field strength affects THz generation and propagation in a $Ti_3C_2T_x$ MXene layer. In panel (a), time-domain THz waveforms for various peak field amplitudes exhibit a clear nonlinear dependence: low-intensity excitation yields weak, structureless signals, while higher intensities produce stronger, spectrally richer fields with sharper temporal features. This nonlinear response originates from enhanced ponderomotive and photothermal forces that accelerate free carriers more efficiently as the field increases. At elevated intensities, additional mechanisms—such as impact ionization, field-assisted tunneling, and $\chi^{(3)}$-mediated interband transitions—further amplify transient currents, broadening the THz spectrum. Panel (b) shows corresponding far-field radiation patterns, where increasing intensity broadens angular emission and introduces multi-lobed structures, attributed to enhanced lateral currents and nonlinear propagation. These results demonstrate that THz generation efficiency and directivity are strongly modulated by laser intensity, although excessive fields may cause carrier saturation, optical damage, or coherence loss. Figure 7 explores the influence of MXene layer thickness on THz emission characteristics. Panel (a) reveals a non-monotonic dependence of the temporal THz waveform on thickness. A 10 nm layer produces sharp, high-frequency oscillations due to quasi-ballistic carrier transport and minimal reabsorption. At 20 nm, the waveform becomes smoother and slightly weaker, reflecting increased electron–phonon scattering and field inhomogeneity, yet maintaining broadband features. For 40 nm thickness, the field shifts toward lower frequencies with reduced fine structure. The low-frequency shift observed at 40 nm thickness originates from two physically distinct mechanisms that become simultaneously relevant when the MXene layer is no longer ultrathin. First, increasing the thickness enhances the effective carrier transit time $\tau_t$ across the layer. Since $\tau_t$ becomes comparable to the sub-picosecond laser-driven current oscillation period, the high-frequency components of the transient current are damped more strongly due to enhanced intralayer scattering and recombination and shifts the spectral maximum toward lower frequencies. Second, the electromagnetic propagation inside the MXene film transitions from the single-interface regime ($d \ll \lambda_{th}z$) to a regime where $d/\lambda_{th}z \gtrsim 0.01$–$0.05$, allowing the formation of weak Fabry–Pérot–type internal reflections. These multiple reflections introduce an additional phase delay that selectively attenuates higher-frequency components due to their larger phase accumulation. Panel (b) shows that thickness also affects spatial emission: thinner layers yield tightly directed, dipole-like radiation, while thicker ones exhibit broader, multi-lobed patterns caused by interference among distributed current sources across the film depth. These interference effects intensify under elliptical polarization, emphasizing the need to balance absorption, dispersion, and coherence for optimized THz emission and directivity.

Material-dependent factors such as conductivity, carrier lifetime, surface termination chemistry, and lattice symmetry varied the temporal characteristics and far-field behavior of THz pulses. Figure 8 indicates the influence of MXene material composition on both the temporal evolution of the emitted THz electric field and it's corresponding far-field radiation pattern under elliptical laser polarization. Specifically, three representative MXene systems—$Ti_3C_2T_x$, $Mo_2TiC_2T_x$, and $Nb_2CT_x$—are investigated to reveal how intrinsic material properties modulate the efficiency, bandwidth, and directivity of THz generation. In panel (a), the normalized electric field is plotted as a function of time for each material. All three waveforms exhibit ultrafast transient characteristics with sub-picosecond features, consistent with photo-induced carrier acceleration dynamics. Notable distinctions arise in the peak amplitudes, oscillation periodicity, and temporal symmetry of the THz waveforms among different MXenes, reflecting variations in their electronic structures, carrier mobility, and interlayer coupling. The $Ti_3C_2T_x$ sample exhibits a moderately peak followed by damped oscillations, indicating efficient initial carrier acceleration but rapid decoherence due to scattering or carrier trapping. In contrast, $Mo_2TiC_2T_x$ produces the highest initial field amplitude and a more symmetric waveform, suggesting enhanced coherence and a sustained photoresponse. This behavior likely originates from the Mo–Ti alloyed sublattice, which provides favorable band alignment and a reduced carrier effective mass, thereby promoting efficient THz emission. The $Nb_2CT_x$ response lies between the two, showing slightly sharper oscillations and higher frequency components, possibly due to niobium's heavier atomic mass and d-electron configuration that enhance intraband transition dynamics and field screening effects.

Subtle time-domain shifts among the materials also indicate variations in group velocity and dispersion of the generated THz pulses, consistent with experimental observations on MXene-dependent THz emission [41]. In panel (b), all three materials display forward-directed radiation, but their far-field patterns differ in beam width and side-lobe structure. $Mo_2TiC_2T_x$ exhibits the most focused and intense lobe along the 0° axis, signifying constructive interference and spatial coherence from a uniform current source. $Ti_3C_2T_x$ shows broader and more asymmetric emission, implying distributed or laterally inhomogeneous carrier transport, while $Nb_2CT_x$ presents a smoother, more symmetric pattern indicative of intermediate coherence. These differences underscore the crucial influence of material composition, dielectric environment, and interfacial charge dynamics on THz emission efficiency and directionality in MXene-based sources.

Figure 9 examines the effect of equilibrium carrier density on the temporal characteristics of THz radiation emitted from a $Ti_3C_2T_x$ MXene layer under femtosecond laser excitation. The time-domain results display oscillatory THz patterns whose amplitude and frequency strongly depend on the carrier concentration. As electron density increases, the peak electric-field amplitude initially rises, reflecting an enhanced nonlinear response dominated by higher-order current terms associated with the second- and third-order susceptibilities. However, beyond a critical density, the waveform becomes smoother and attenuated due to increased carrier–carrier scattering and screening, which damp transient current oscillations. This behavior highlights the dual role of carrier density: moderate levels enhance THz emission through stronger photoinduced currents, whereas excessive densities induce intraband scattering that suppresses the net modulation responsible for radiation. The observed phase shifts and slight temporal asymmetry at higher densities indicate augmented Drude-like conductivity and intensified nonlinear polarization effects. The dependence of THz radiation on carrier density is consistent with experimental findings on 2D MXenes [40].

The temporal evolution of the normalized THz electric field emitted from a $Ti_3C_2T_x$ MXene layer, with variations induced by differing temperature conditions, is delineated in Fig 10. Available THz–TDS and transport data for $Ti_3C_2T_x$ [30, 32] indicate that near room temperature the linear approximation is adequate. Therefore, in the present simulations the pragmatic parametrization $\gamma(T) = \gamma_0 + \alpha T$ is adopted, where $\gamma_0$ collects the temperature-independent channels and $\alpha$ encapsulates the slope from electron–phonon scattering in the near-room-temperature regime. One notes that $\alpha$ is an effective parameter that depends on both intrinsic phonon spectra and extrinsic surface terminations (–O, –OH, –F) which modify scattering phase space [40]. The underlying mechanism is traced to the temperature dependence of the carrier momentum relaxation rate, expressed through the damping coefficient $\gamma$ in the Drude model. At higher temperatures, increased phonon populations and stronger electron–phonon scattering broadens the carrier momentum distribution and accelerate energy dissipation, effectively increasing $\gamma(T)$. This enhanced damping shortens the lifetime of photoinduced currents and reduces their coherence, thereby attenuating the radiated THz field. The resulting suppression of higher-frequency components in the waveform further reflects this thermally induced broadening of the carrier response. The relationship between temperature and be modeled as a thermally activated process, where T in the low-temperature regime transitions to a stronger dependence at elevated temperatures due to multi-phonon interactions. This is consistent with the Drude model adapted for two-dimensional materials. In addition to phonon interactions, the temperature rise also modulate carrier scattering via surface or chemical termination groups (e.g., –O, –OH, –F) commonly present on MXene surfaces. These groups act as dynamic scattering centers whose activity increases with thermal energy, introducing an additional dissipation channel. Thus, temperature not only governs intrinsic electron–phonon interactions but also activates extrinsic damping mechanisms tied to surface chemistry, offering a multifaceted control over the THz emission dynamics in MXene-based systems. The effect of temperature on THz radiation from a $Ti_3C_2T_x$ MXene layer is provided by experimental findings demonstrated in Ref. [32].

Figure 11 reveals the impact of pump laser wavelength on the temporal structure of THz radiation generated from a $Ti_3C_2T_x$ MXene layer, highlighting the role of surface plasmon dynamics in the nonlinear emission process. The time-resolved electric field traces, normalized to the peak input field, demonstrate a marked dependence on excitation wavelength, with both the waveform complexity and amplitude exhibiting nonmonotonic variation across the 400–1600 nm range. At shorter wavelengths, the THz signal displays rich temporal modulations and high-frequency oscillations, indicative of interband transitions and rapid polarization dynamics. As the wavelength increases, the waveform becomes more coherent and intense, suggesting optimal coupling between the incident optical field and resonant plasmon modes supported by the MXene layer. This enhancement is due to near-field excitation of surface plasmons, whose resonance condition depends sensitively on both material permittivity and pump frequency. At even longer wavelengths, the field exhibits increasingly irregular and broadband oscillations. This behavior arises from broadband plasmonic excitation with diminished field localization and reduced overlap with nonlinear charge carrier dynamics. As the wavelength modulates, a shift in the oscillatory frequency of the emitted THz field becomes evident, reflecting the resonant coupling between the incident photon energy and the electronic transitions within the MXene material. This frequency adjustment arises from the inverse relationship between wavelength and photon energy, which alters the efficiency of nonlinear processes—such as second- and third-order susceptibilities—driving the THz generation. The results in the impact of pump laser wavelength on the temporal structure of THz radiation generated from a $Ti_3C_2T_x$ MXene layer, are consistent with experimental findings [32].

The temporal evolution of the normalized THz electric field emitted from a $Ti_3C_2T_x$ MXene layer, with variations induced by the material properties of the underlying substrate, is explored in Fig 12. The normalized electric field traces reveal that the choice of underlying substrate plays a significant role in shaping the time-domain characteristics of the emitted THz pulse. Differences in waveform morphology, peak amplitude, and spectral richness point to substrate-dependent modifications in both electromagnetic boundary conditions and charge carrier dynamics at the interface. The substrate affects the THz response primarily through its dielectric constant, optical absorption, and phonon-polariton resonances, all of which influence how the radiated field propagates and interferes near the surface. The temporal dynamics indicate that substrates with higher dielectric constants or metallic characteristics facilitate stronger plasmonic resonances, leading to enhanced peak amplitudes and modified oscillation frequencies. This enhancement arises from the excitation of surface plasmon polaritons (SPPs) at the interface, where the collective oscillation of free electrons in the substrate couples with the incident optical field, boosting the nonlinear polarization terms within the MXene layer. Conversely, substrates with lower dielectric constants or insulating properties exhibit reduced plasmonic coupling, resulting in diminished THz output and increased damping due to weaker field confinement. For instance, substrates supporting surface phonon polaritons induce hybrid plasmon–phonon modes, thereby altering the temporal coherence and decay rate of the emitted THz field. These interactions also modify the effective local field acting on photoexcited carriers, thereby modulating the amplitude and phase of the THz waveform.

The temporal evolution of the normalized THz electric field component emitted from a $\mathrm{Ti_3C_2T_x}$ MXene layer with varying full width at half-maximum (FWHM) durations is illustrated in Fig 13. As the FWHM increases, the THz waveform becomes progressively more structured and temporally extended, suggesting enhanced coherence and stronger spectral mixing. This behavior is attributed to the increased interaction time between the optical field and the charge carriers, which facilitates the buildup of collective oscillations and more pronounced transient photocurrents. In this regime, the ponderomotive force, and nonlinear polarization terms remain sufficiently to drive resonant electron motion, leading to the generation of temporally rich and high-frequency field components. At shorter durations, the waveform exhibits sharper transients with broader spectral features but reduced overall amplitude. This is indicative of impulsive excitation, where the ultrashort pulse duration limits energy transfer efficiency and confines carrier excitation to sub-picosecond timescales. While such pulses can access high-frequency components due to their broadband nature, the reduced interaction time results in weaker nonlinear buildup and less effective charge acceleration. In contrast, for the longest pulse duration, a reduction in temporal fine structure is observed, along with noticeable attenuation of peak field amplitude. This transition marks a regime where optical excitation becomes increasingly quasi-continuous, diminishing the effectiveness of ultrafast nonlinear processes. The extended pulse envelope allows for greater energy dissipation through electron–phonon scattering and other relaxation channels, leading to loss of coherence and suppression of higher-frequency components in the emitted THz signal.

Figure 14 presents the normalized amplitude spectrum of the emitted THz radiation, comparing the cases with and without the inclusion of nonlinear terms modelling bound electrons (i.e., $\mathbf{P(r,t)} = \epsilon_0\left[\chi^{(2)}\mathbf{E(r,t)E(r,t)} + \chi^{(3)}\mathbf{E(r,t)E(r,t)E(r,t)}\right]$) in the governing equations of the light-matter interaction. The frequency-resolved results clearly demonstrate the decisive influence of nonlinear optical effects—originating from field-induced carrier acceleration, interband transitions, and higher-order polarization—on both the amplitude and bandwidth of THz emission. These nonlinearities arise from multiphoton absorption, ponderomotive forces, and anharmonic charge motion within the MXene layer or its substrate. When nonlinear terms are incorporated, the spectrum broadens significantly across 0–5 THz and exhibits enhanced amplitudes, along with distinct resonant peaks near 0.4, 2.2, and 4.6 THz. These features indicate complex temporal dynamics, nonlinear phase modulation, and wave-mixing effects. In contrast, the purely linear case yields a weaker, low-frequency spectrum (< 2 THz) with monotonic decay and no higher harmonics, reflecting limited current modulation and absence of nonlinear coupling. This comparison underscores the essential role of nonlinear polarization in enabling high-intensity, broadband THz emission. The bound-electron response, modeled through nonlinear susceptibilities $\chi^{(2)}$and $\chi^{(3)}$, represents virtual transitions and anharmonic oscillations of tightly bound electrons, distinct from the free-carrier Drude contribution. In MXenes, these nonlinearities facilitate optical rectification and second-harmonic generation—key mechanisms for broadband THz production [25]. Experimental measurements reporting $\chi^{(3)} \approx 10^{-10}$ esuin $Ti_3C_2$ MXenes under near-IR excitation confirm that surface terminations (–O, –OH, –F) enhance bound-electron polarizability and contribute strongly to THz generation in the 0.1–5 THz range [40, 41]. The peaks at 0.4–5 THz correspond to nonlinear mixing between the drive frequency and the intrinsic plasmon frequency $\omega_p \approx 2 - 5\, THz$ for $Ti_3C_2T_x$. The required field $E_0 \approx 10^7$ V/m corresponds to intensities $\approx 10^{12}$–$10^{13}$ W/cm², below the reported optical-damage threshold of MXenes (~$10^{14}$ W/cm²). To provide a direct comparison with established THz-emission technologies, the efficiency of THz generation in MXenes, derived from FDTD simulations, ranges from η ~ 10⁻3–10⁻4 for single-layer MXenes (peak fields ~$10^6$–$10^7$ V/m at input intensities ~$10^{13}$ W/cm²) to potentially η ~ $10^{-2}$ for optimized multilayer stacks, comparable to graphene (η ~ $10^{-5}$–$10^{-3}$) [26]. Compared to air-plasma generation (η ~ $10^{-5}$–$10^{-4}$, high energy requirements (>mJ) and bulky setups) [42], MXenes offer compact, solid-state alternatives. For context, the simulated MXene emission (0.1–1 MV/m peak field, 0.3–5 THz bandwidth, and η ≈ $10^{-3}$–$10^{-4}$) is consistent with the performance range of conventional THz sources. $LiNbO_3$ typically reaches η ≈ $10^{-4}$–$10^{-3}$ but with narrower spectra (< 2.5 THz), while ZnTe operates at η ≈ $10^{-6}$–$10^{-4}$ with bandwidths up to ~3 THz. InAs and InAs-nanowire emitters provide broader spectra (3–4 THz) but modest efficiencies (~$10^{-5}$–$10^{-4}$). The MXene results therefore fall within the physically expected regime and compare favorably with broadband semiconductor emitters despite the ultrathin (20 nm) geometry [43- 45].

The presented model includes dominant intraband nonlinearities but does not explicitly treat interband transitions, hot-carrier diffusion, or phonon-assisted scattering beyond. These effects require quantum-kinetic or ab-initio approaches and are identified as essential future extensions.

**Conclusion**

In the present work, the potential of two-dimensional MXene layers as active platforms for THz wave generation under femtosecond laser excitation has been rigorously investigated through a synergistic combination of HD Drude theory and finite-difference time-domain (FDTD) simulations. By incorporating nonlinear carrier dynamics, plasmonic resonances, and light–matter coupling at ultrafast timescales, a detailed picture of THz radiation mechanisms in MXenes has been constructed. It has been demonstrated that the generation of broadband, high-intensity THz pulses in MXenes arises from a complex interplay of ponderomotive and photothermal forces, nonlinear polarization currents, and resonant plasmonic excitations. The temporal and spatial characteristics of the emitted THz field were found to be highly tunable through laser parameters such as incidence angle, polarization state, intensity, and pulse duration. Moreover, the intrinsic properties of the MXene layer, including its thickness, equilibrium carrier density, composition, and temperature, were shown to exert influence on both the efficiency and spectral content of the emitted radiation. Notably, the inclusion of nonlinear source terms in the governing equations has been shown to significantly enhance THz field strength and extend the spectral bandwidth, yielding distinct harmonic peaks attributable to multi-photon absorption and wave-mixing phenomena. This study provides, for the first time, a comprehensive theoretical framework that treats MXenes not merely as passive absorbers but as dynamically active THz emitters capable of generating directionally biased THz radiation with temporal phase relationships preserved at the sub-cycle level. The demonstrated sensitivity of the emission process to a multitude of physical and geometrical parameters opens a pathway toward the rational design of ultrathin, chip-integrable THz sources. These findings are expected to have far-reaching implications for the development of next-generation THz photonic systems, spanning ultrafast spectroscopy, high-resolution imaging, and nanoscale wireless communication. Future directions include the experimental validation of the proposed mechanisms, the exploration of heterostructures combining MXenes with other 2D materials to tailor plasmonic and nonlinear responses, and the integration of strain-engineered or thermally modulated devices for real-time tunable THz applications. The simulated THz pulses exhibit peak fields of 0.1–1 MV/m at 1 mm distance, spectral bandwidth up to 5 THz, and conversion efficiency $\eta \approx 10^{-3}$–$10^{-4}$ for single-layer $Ti_3C_2T_x$ under $10^{12}$–$10^{13}$ W/cm² excitation. Unlike an ideal 2DEG, MXenes possess surface terminations that modify carrier scattering through localized dipoles and phonon–surface interactions. Additionally, specific THz-TDS validation protocols are outlined: measurement of pump-dependent THz waveform, thickness-dependent phase delay, and temperature-dependent damping, all directly comparable to presented predictions.

## Acknowledgements

Authors are thankful to Iran National Science Foundation (INSF) for the financial support provided under the grant number 4030055.

## Declaration of competing interest

The authors declare that they have no known competing financial interests or personal relationships that could have appeared to influence the work reported in this paper.

## Data availability

Data will be made available on request.

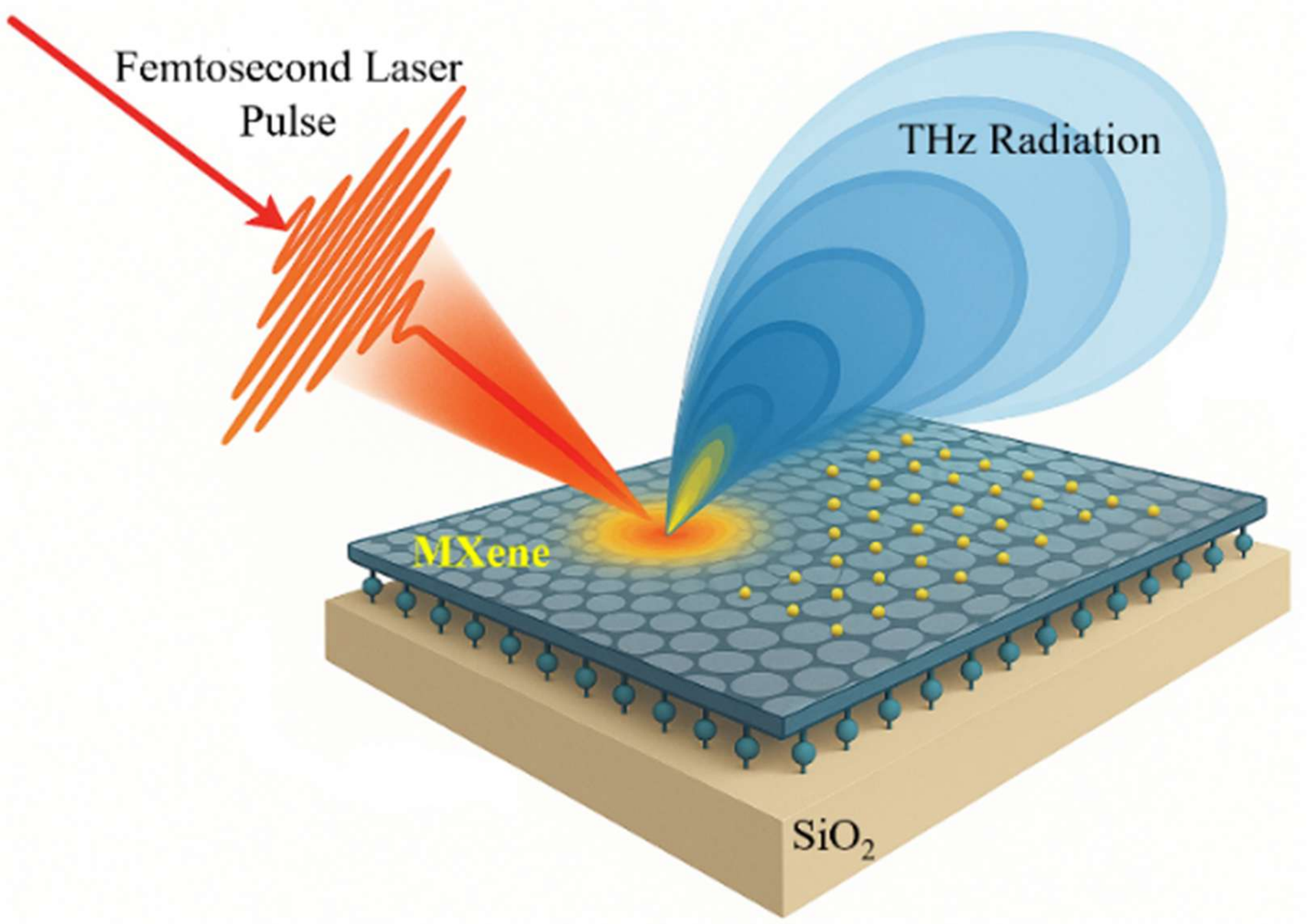

FIG 1. Schematic representation of THz wave generation from a two-dimensional MXene layer upon femtosecond laser excitation. The ultrafast laser pulse drives nonlinear charge dynamics in the MXene, resulting in the emission of broadband THz radiation into the far field.

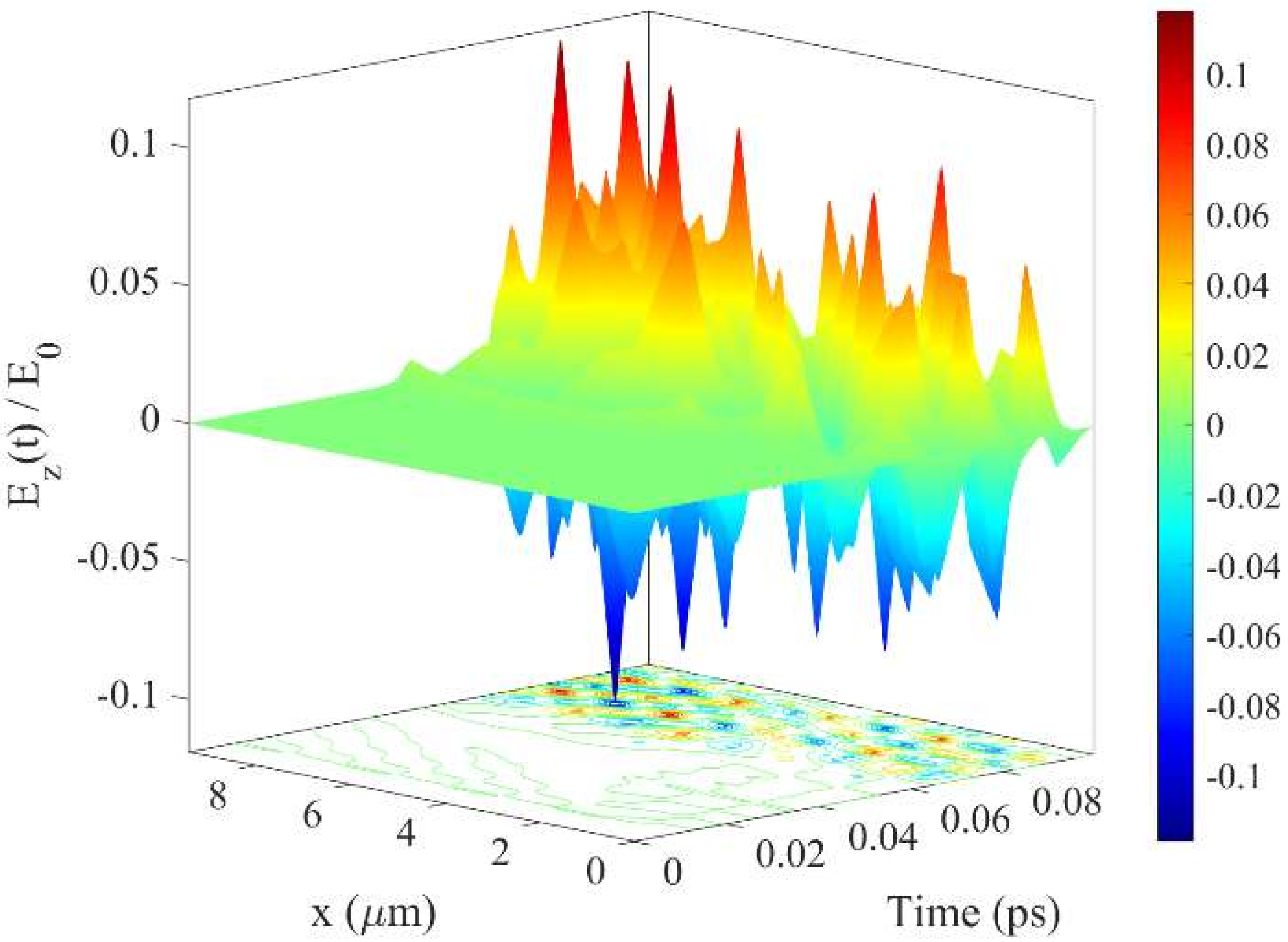

FIG 2. Three-dimensional plot showing the spatial and temporal evolution of $E_z$ with contour in the MXene layer.

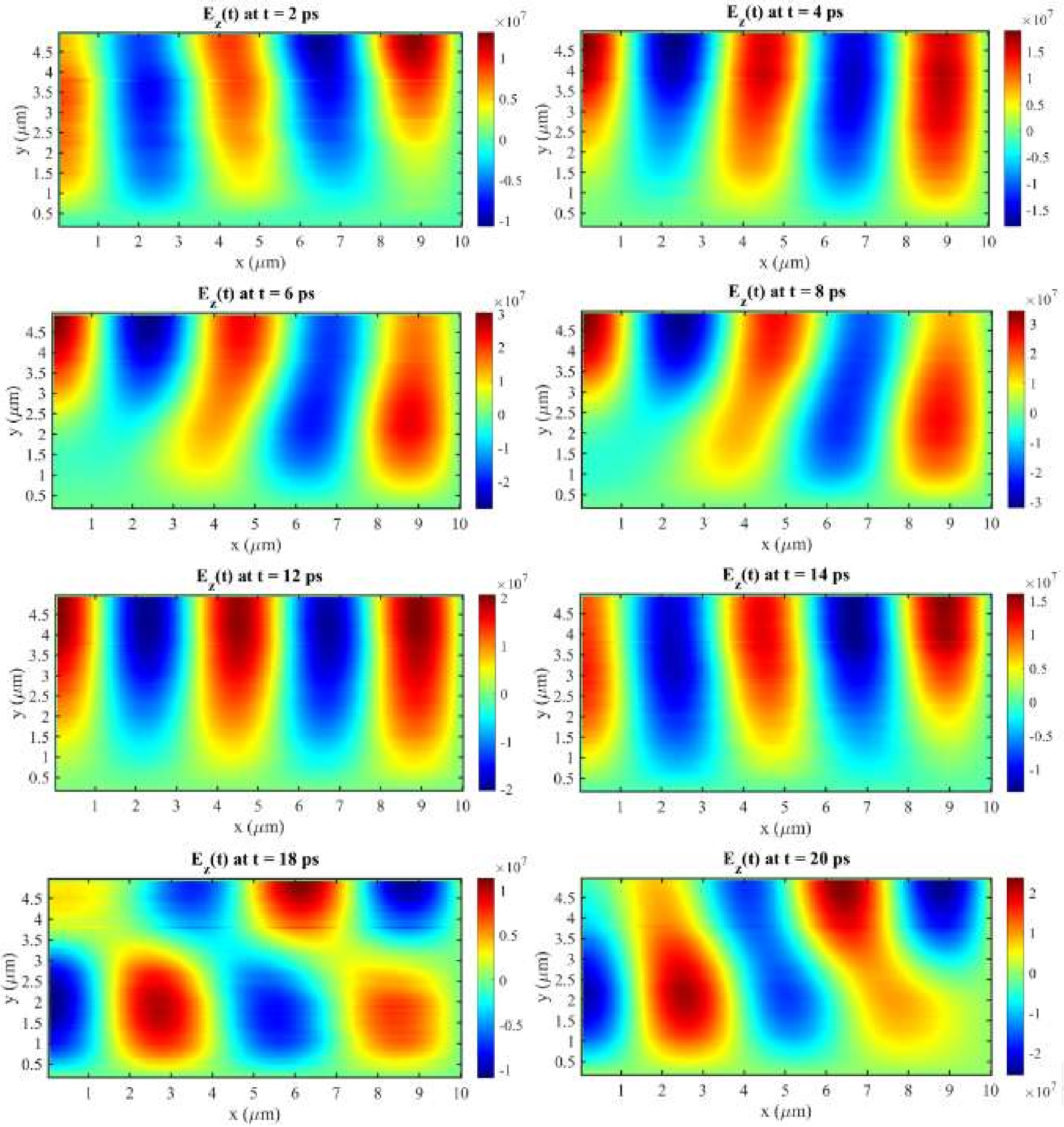

FIG 3. Time-resolved two-dimensional snapshots of the electric field component $E_z$, illustrating the spatiotemporal propagation and modulation of THz waves in the $x - y$ plane over a 20 ps interval.

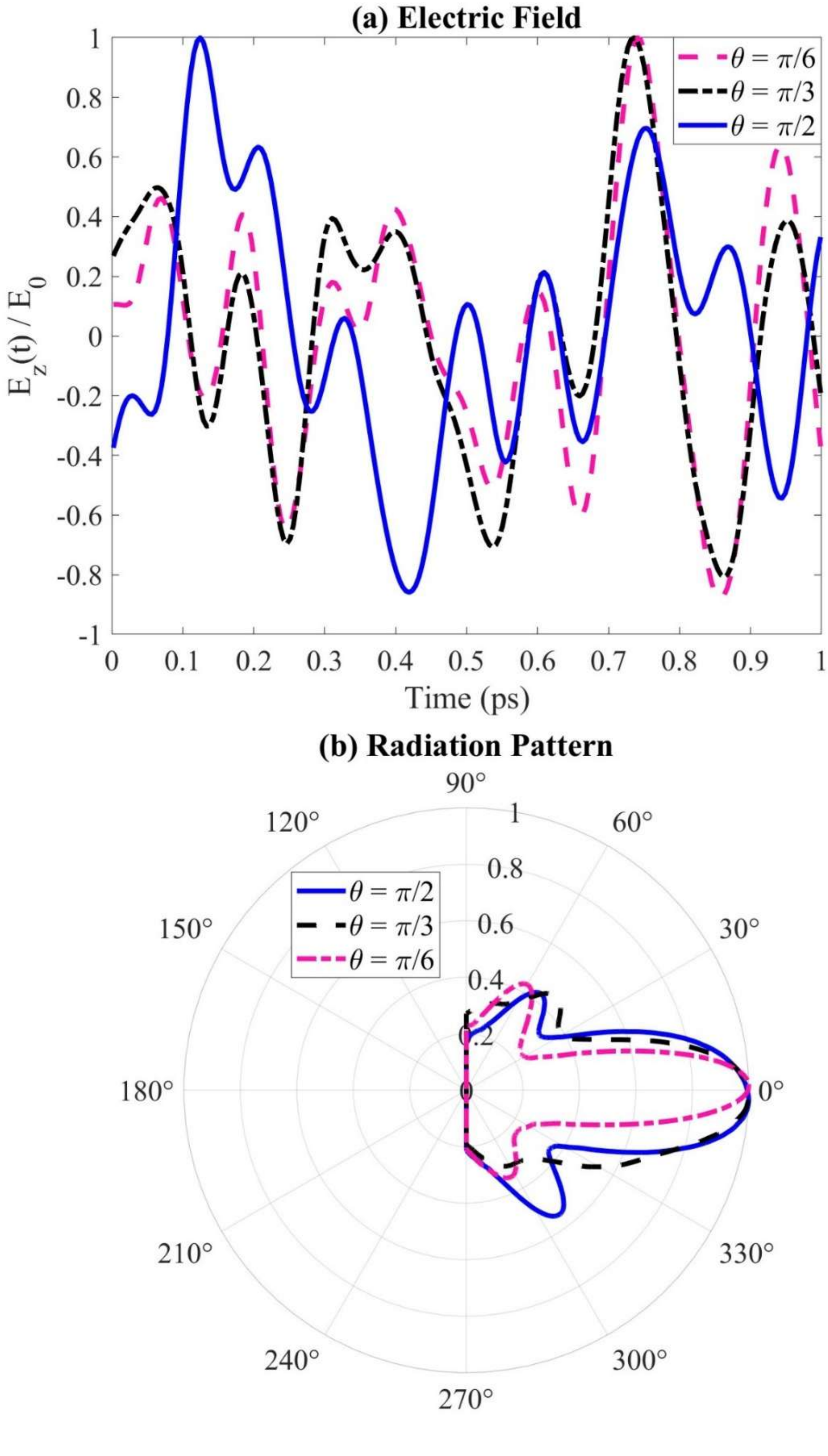

FIG 4. (a) Temporal profiles of the normalized THz electric field generated under different laser incidence angles. (b) Corresponding far-field radiation patterns, highlighting the effect of oblique excitation and elliptical polarization on emission directivity.

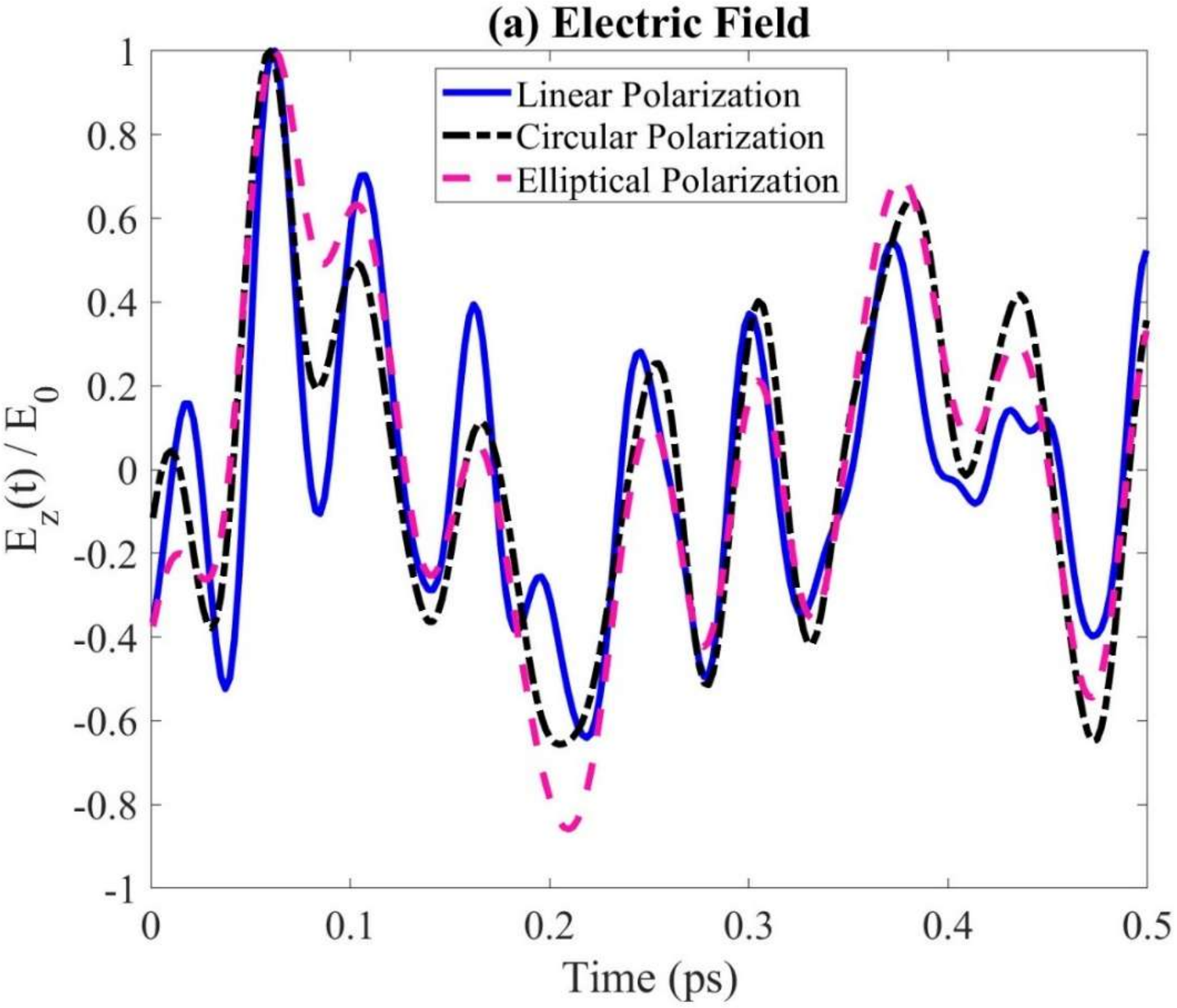

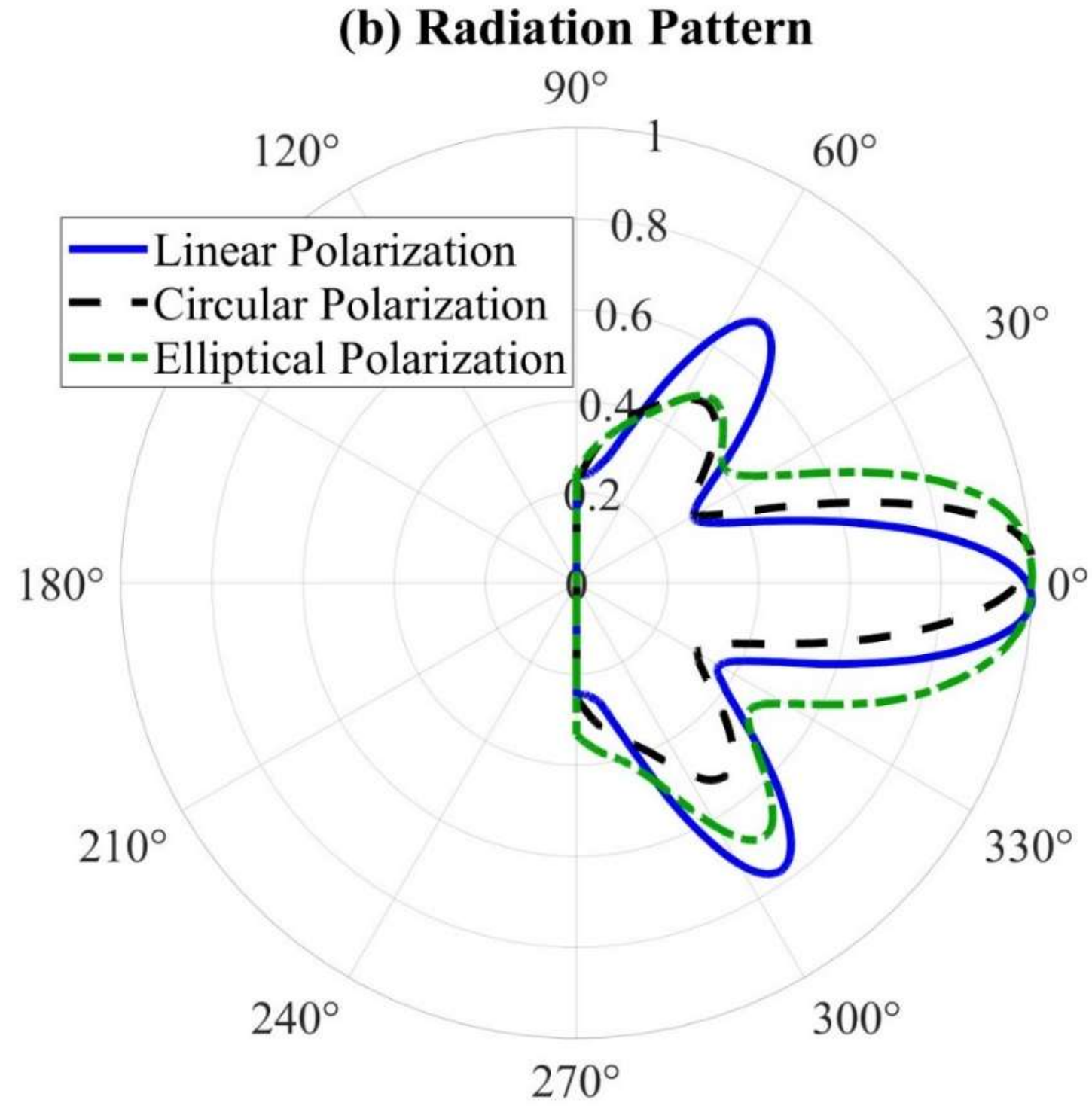

FIG. 5 (a) Effect of incident laser polarization (linear, circular, elliptical) on the temporal characteristics of the emitted THz field. (b) Associated far-field radiation profiles demonstrating polarization-dependent symmetry and directionality.

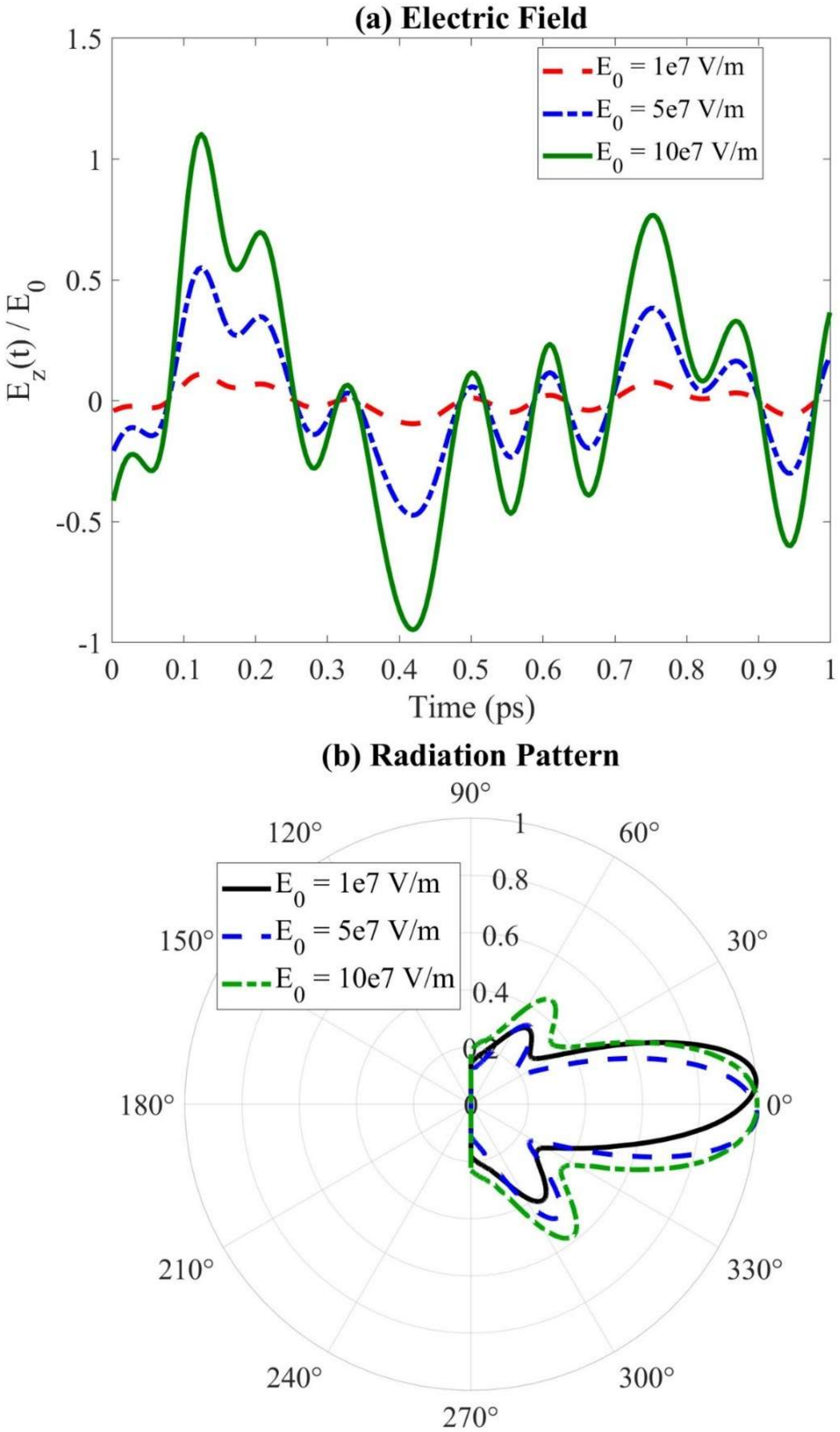

FIG 6. (a) Dependence of THz waveform shape and amplitude on the peak intensity of the incident laser pulse. (b) Far-field radiation patterns corresponding to different intensities, revealing nonlinear enhancement and angular broadening of the emitted field.

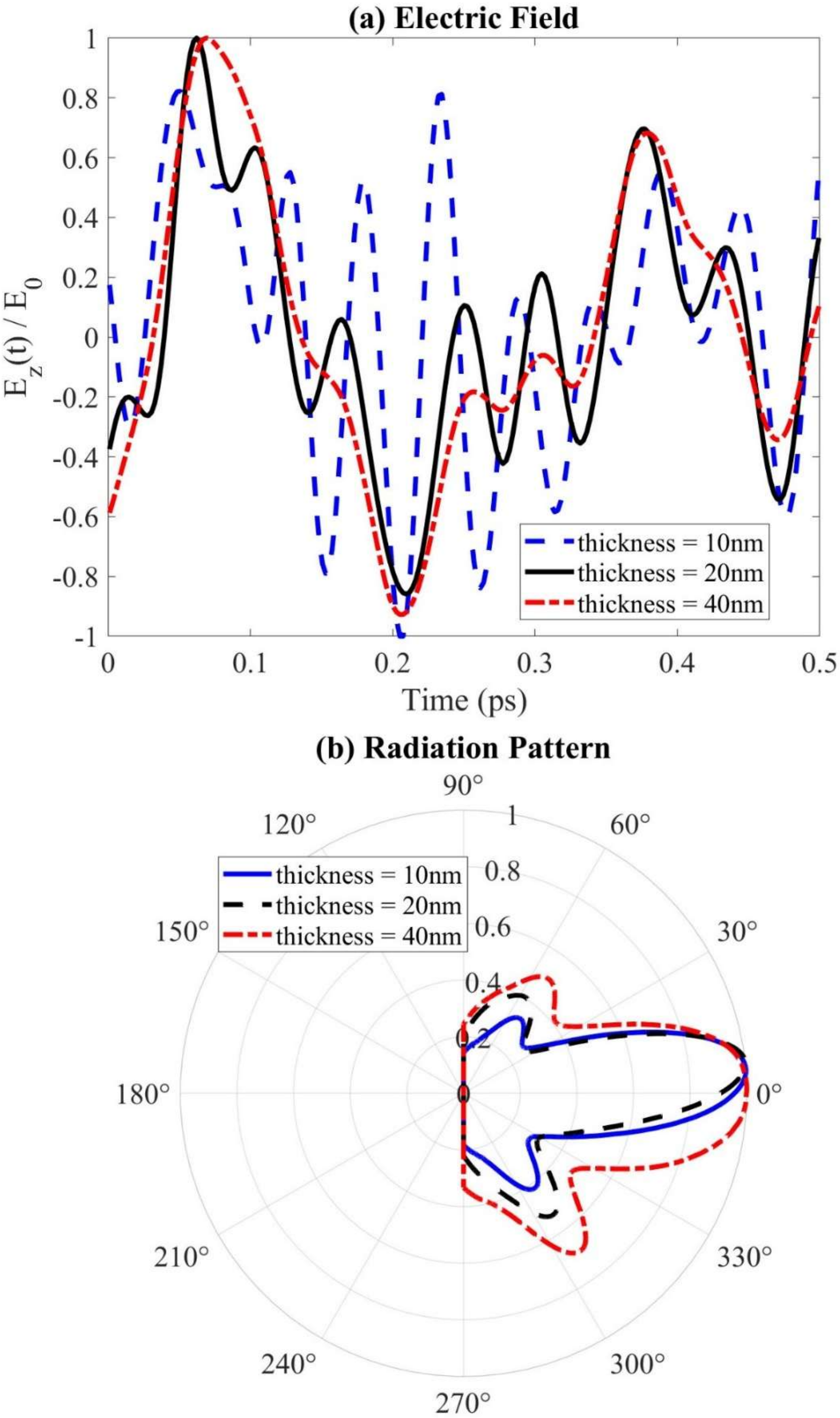

FIG 7. (a) Influence of MXene layer thickness on the temporal profile of the generated THz signal. (b) Far-field radiation patterns for each thickness, showing the role of propagation distance and absorption in shaping emission characteristics.

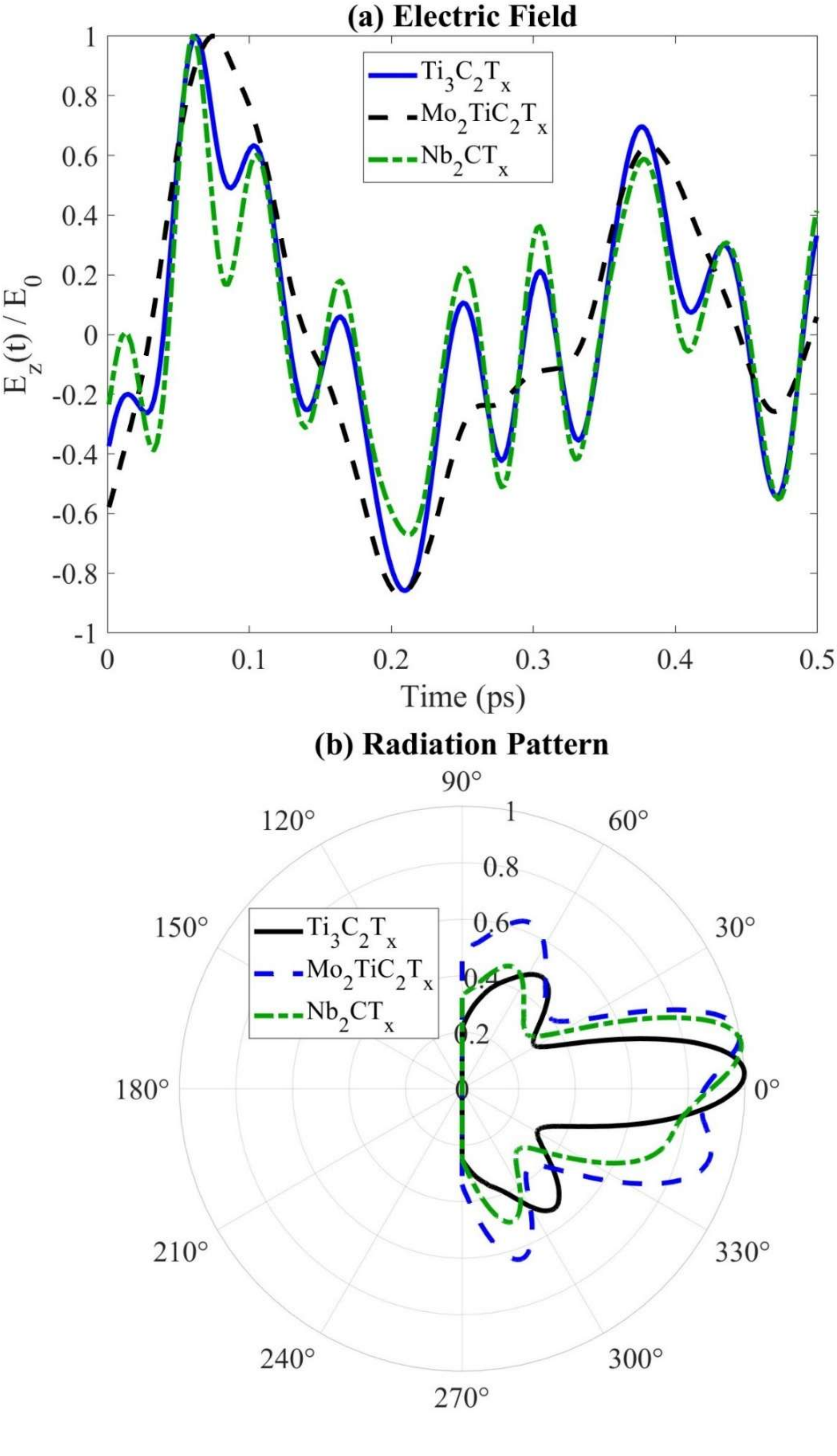

FIG 8. (a) Temporal evolution of the normalized THz electric field emitted from different MXene compositions. (b) Corresponding far-field patterns, demonstrating how material-dependent electronic structure modulates emission strength and beam profile.

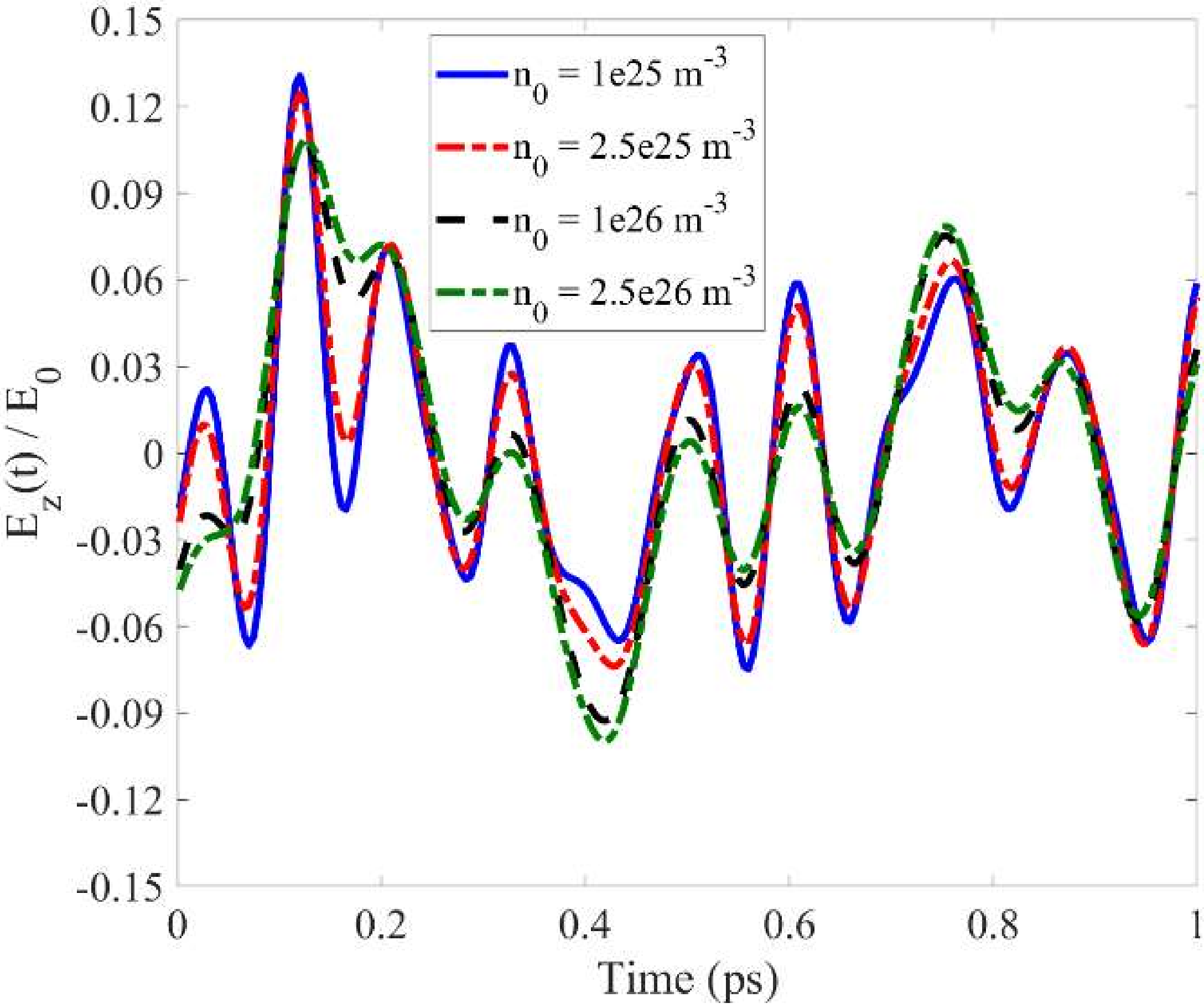

FIG 9. Variation of the emitted THz field waveform with equilibrium carrier density in the MXene layer, showing nonlinear modulation of amplitude and frequency content due to plasma screening effects.

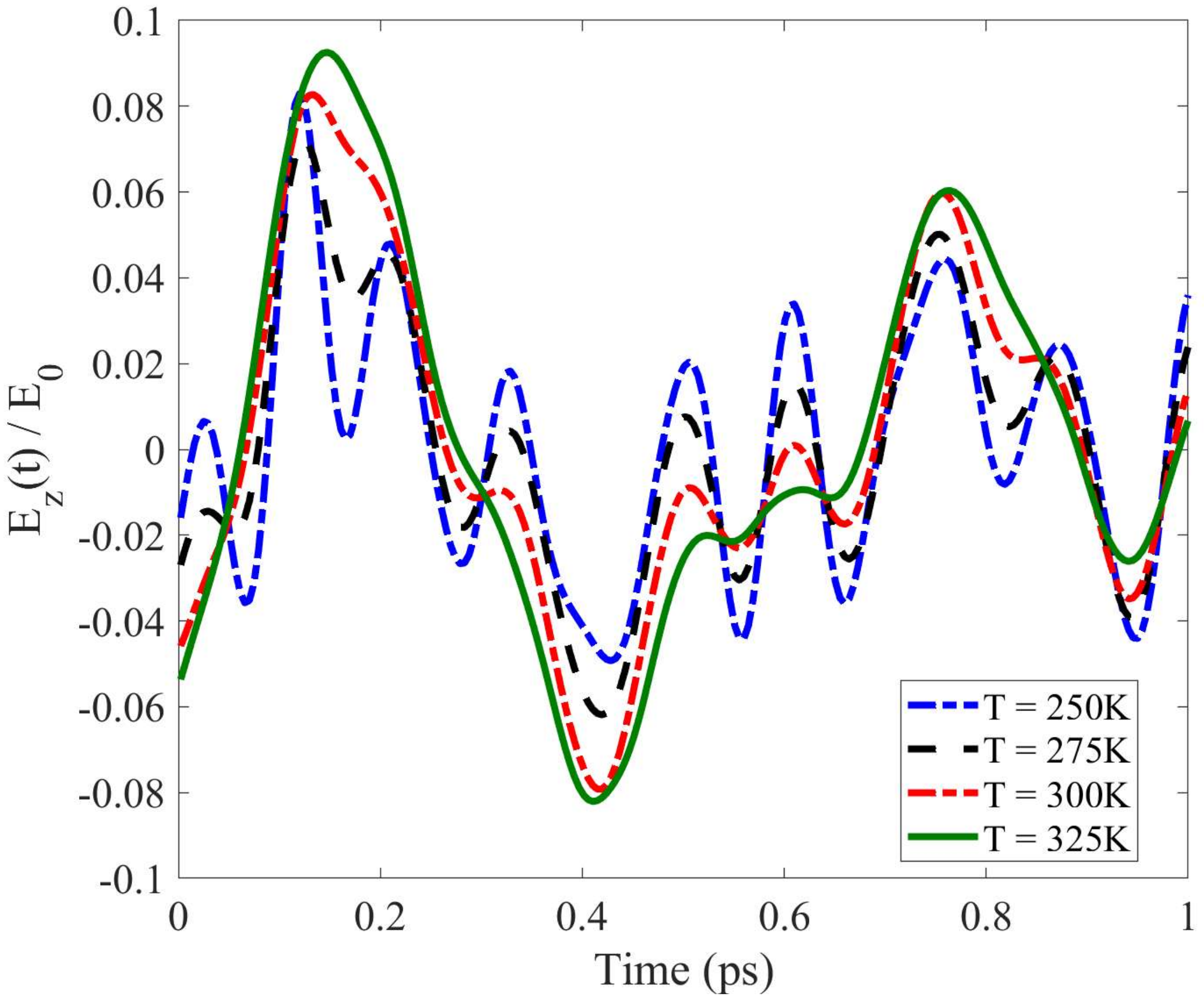

FIG 10. Effect of lattice temperature on the temporal structure of the THz waveform, attributed to thermally activated carrier scattering and enhanced damping in the MXene layer.

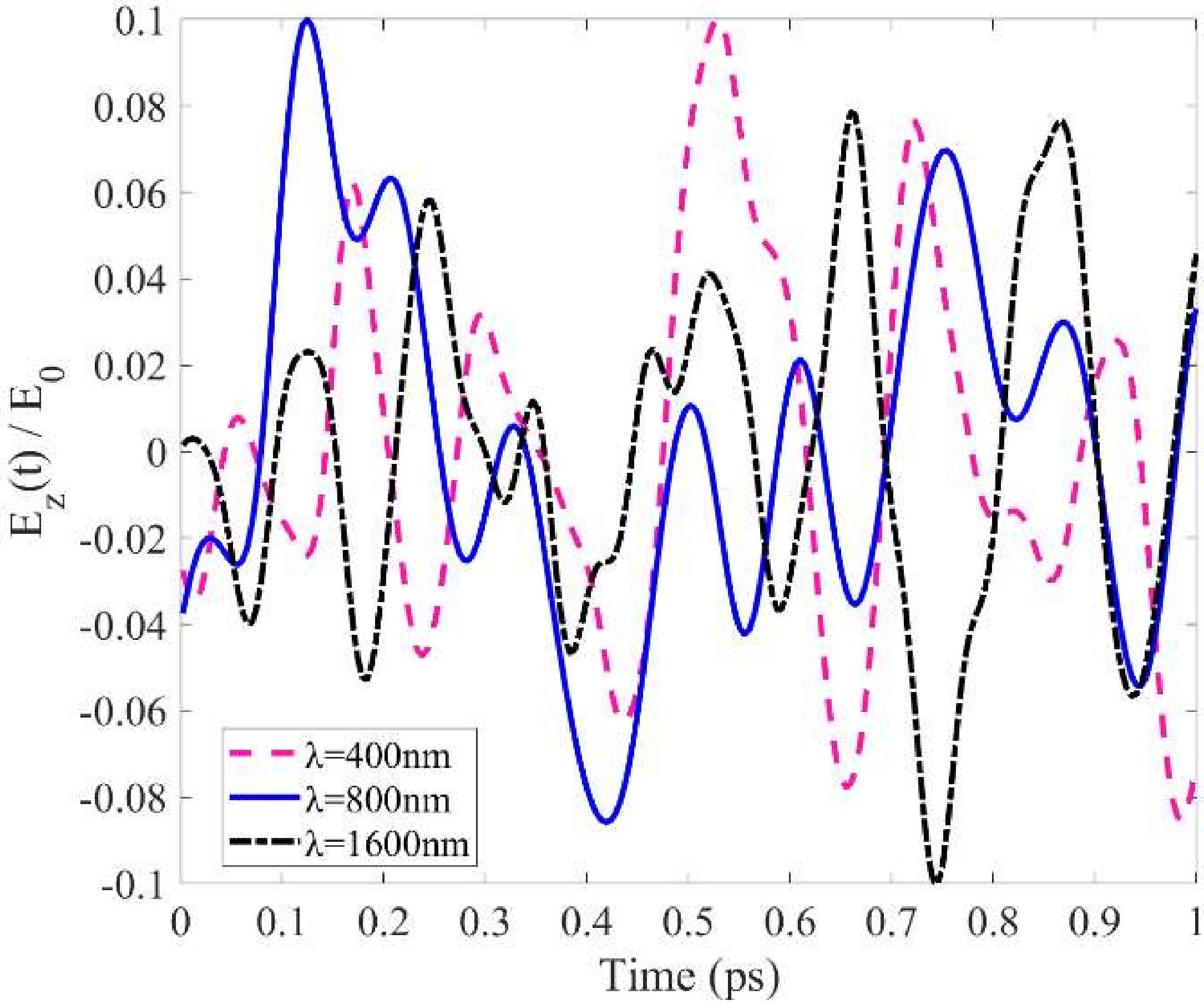

FIG 11. Temporal evolution of the THz electric field for varying pump laser wavelengths, revealing the role of plasmonic resonance and photon energy in shaping emission dynamics.

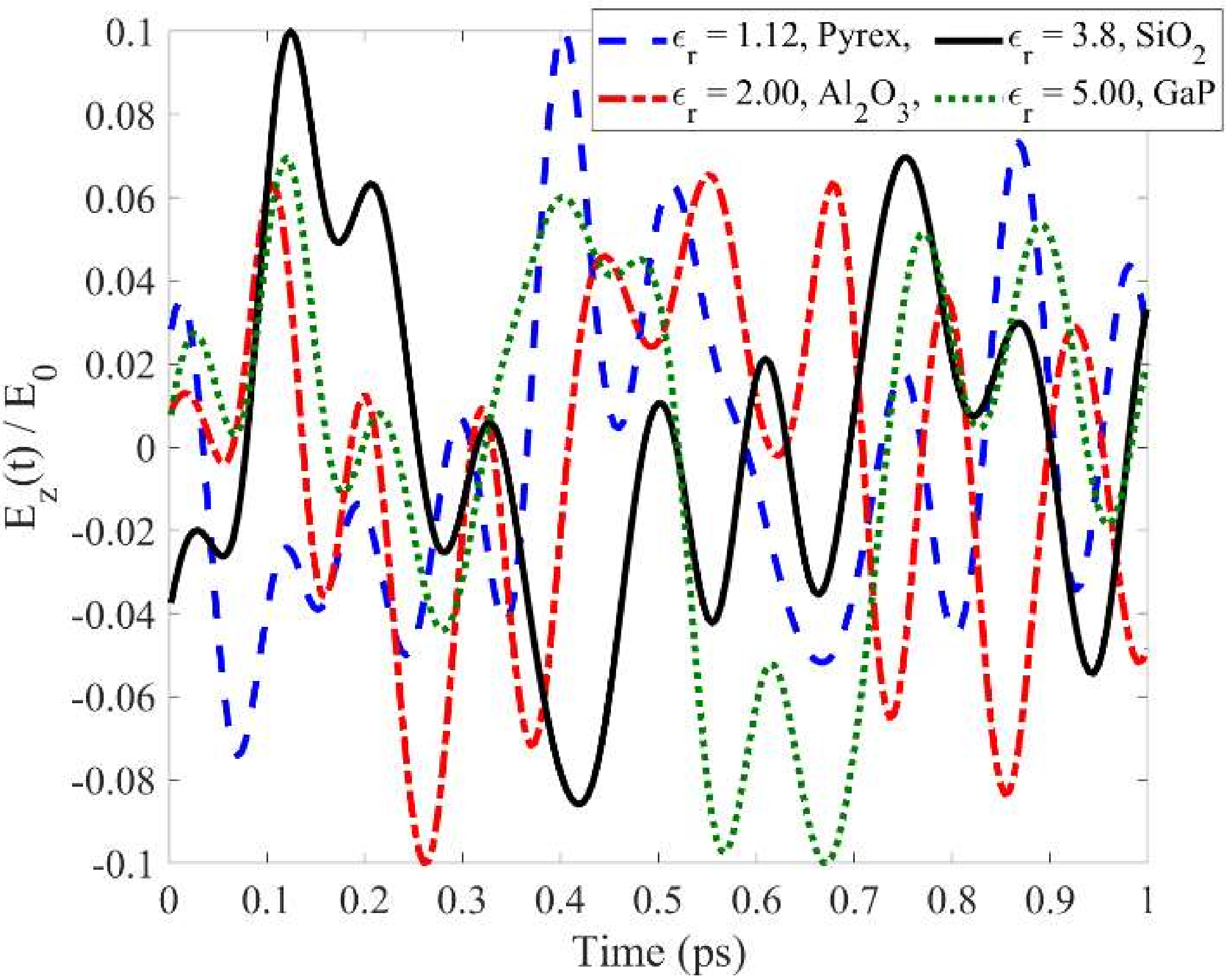

FIG 12. Influence of substrate material on the THz waveform emitted from the MXene layer, demonstrating substrate-mediated modifications in charge dynamics and interfacial field confinement.

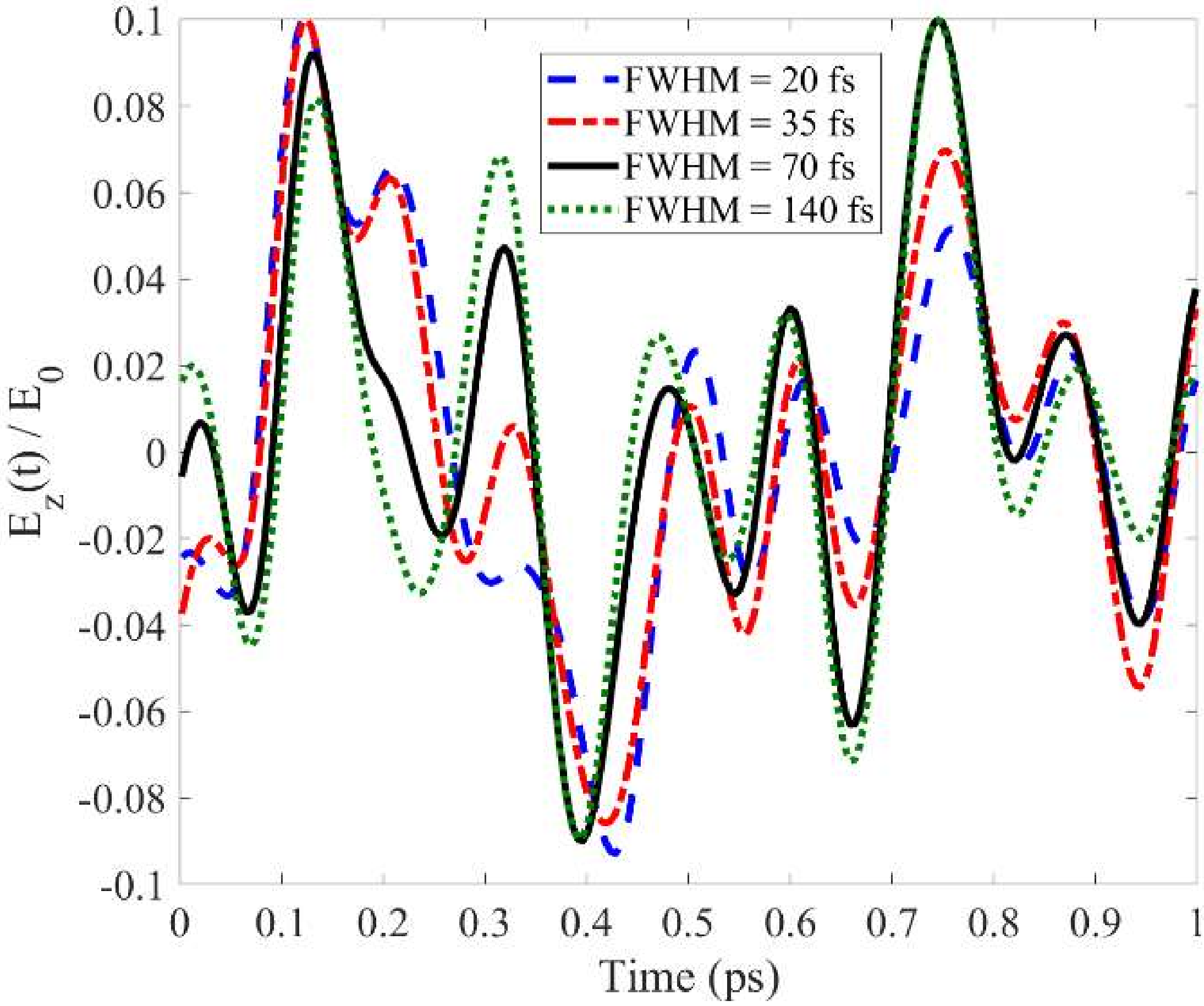

FIG 13. Impact of laser pulse duration (FWHM) on the temporal characteristics of the THz field, illustrating the interplay between interaction time and nonlinear carrier acceleration.

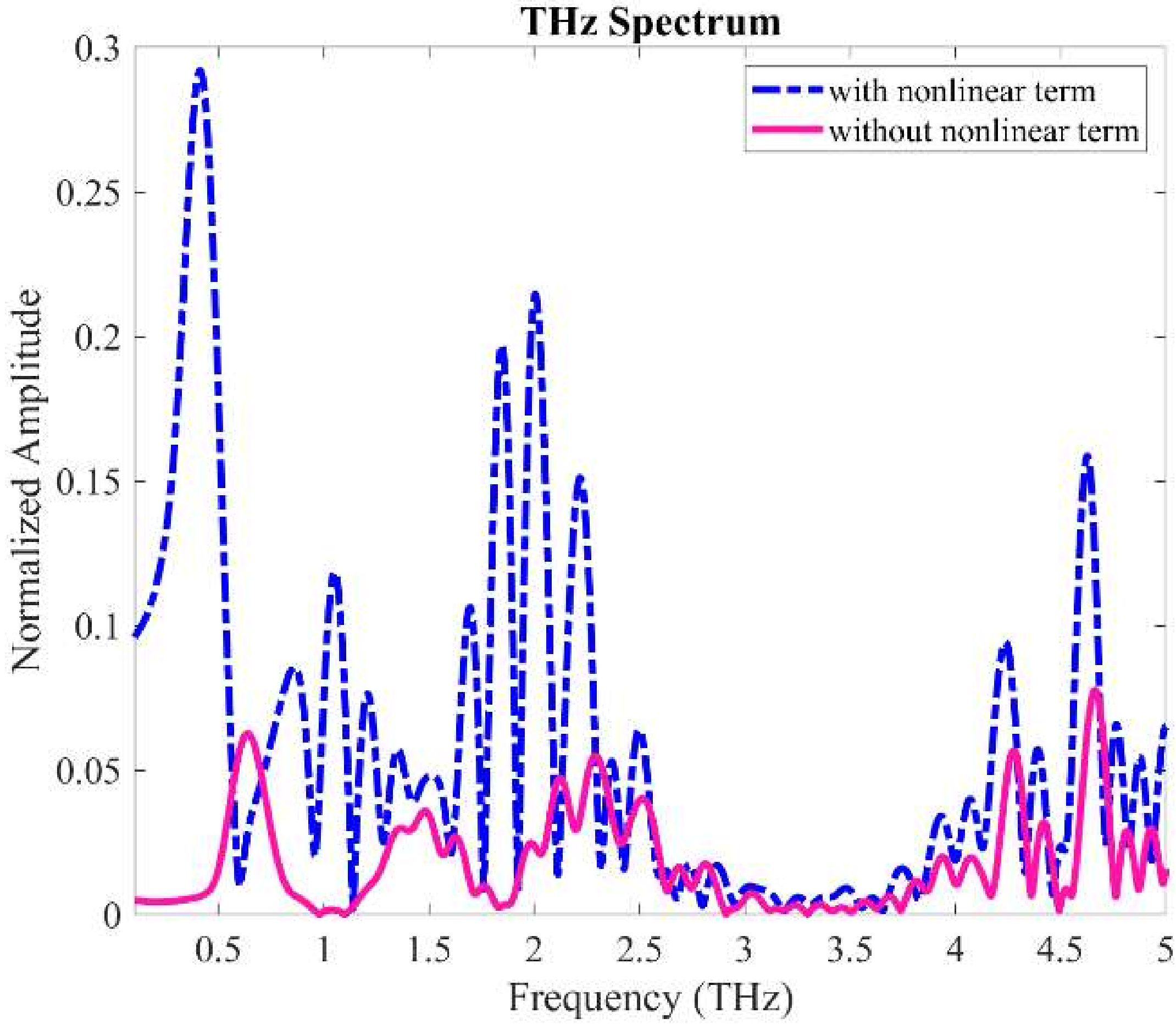

FIG 14. Normalized THz spectral amplitude as a function of frequency, comparing the cases with and without nonlinear polarization $\mathbf{P} = \epsilon_0[\chi^{(2)}\mathbf{EE} + \chi^{(3)}\mathbf{EEE}]$. Spectral broadening and harmonic peaks confirm the critical role of nonlinearities in efficient THz generation.